\documentclass[aps,pre,preprint]{revtex4-1}
\usepackage[caption=false]{subfig}
\usepackage{textcomp}
\usepackage[latin1]{inputenc}
\usepackage{graphicx}
\usepackage{psfrag}
\usepackage{url}
\usepackage{color}
\usepackage{amsmath}
\usepackage{amssymb}
\usepackage{tabularx} 
\usepackage{array}
\usepackage{varwidth}
\usepackage{multirow}
\usepackage{dcolumn}
\usepackage{tipa}
\usepackage{bm}
\usepackage[english]{babel}

\normalfont
\usepackage{booktabs}
\usepackage[bottom]{footmisc}
\newlength\mylength
\begin{document}
\title[Metric deformation method in 3D]{Metric deformation and boundary value problems in 3D}

\author{Subhasis Panda$^{1,}$\footnote{subhasis@cts.iitkgp.ernet.in}}
\author{S. Pratik Khastgir$^{2,1,}$\footnote{pratik@phy.iitkgp.ernet.in}}

\affiliation{$^1$Centre for Theoretical Studies, I.I.T. Kharagpur, 721302, India}
\affiliation{$^2$Department of Physics and Meteorology, I.I.T. Kharagpur, 721302, India}
  
\begin{abstract}
A novel perturbative method, proposed by Panda {\it et al.}
\cite{p3ref4} to solve the Helmholtz equation in two dimensions, is
extended to three dimensions for general boundary surfaces. Although a
few numerical works are available in the literature for specific
domains in three dimensions such a general analytical prescription is
presented for the first time. An appropriate transformation is used to
get rid of the asymmetries in the domain boundary by mapping the
boundary into an equivalent sphere with a deformed interior
metric. The deformed metric produces new source terms in the original
homogeneous equation. A deformation parameter measuring the deviation
of the boundary from a spherical one is introduced as a perturbative
parameter. With the help of standard Rayleigh-Schr{\"o}dinger
perturbative technique the transformed equation is solved and the
general solution is written down in a closed form at each order of
perturbation. The solutions are boundary condition free and which make
them widely applicable for various situations. Once the boundary
conditions are applied to these general solutions the eigenvalues and
the wavefunctions are obtained order by order. The efficacy of the
method has been tested by comparing the analytic values against the
numerical ones for three dimensional enclosures of various shapes. The
method seems to work quite well for these shapes for both, Dirichlet
as well as Neumann boundary conditions. The usage of spherical
harmonics to express the asymmetries in the boundary surfaces helps us
to consider a wide class of domains in three dimensions and also
their fast convergence guarantees the convergence of the perturbative
series for the energy. Direct applications of this method can be found in
the field of quantum dots, nuclear physics, acoustical and
electromagnetic cavities.
\end{abstract}

\maketitle

\section{Introduction}
The three dimensional Helmholtz equation is encountered frequently by
physicist and engineers in different areas $-$ like the eigenanalysis
in acoustic and electromagnetic cavities, transmission of acoustic
waves through ducts and in quantum mechanics. The main concern to
pursue these problems is to calculate the eigenspectrum of the linear
homogeneous Helmholtz equation for different boundary conditions and
geometries. In the acoustic wave motion, for the sinusoidal variation
of the sound pressure with time, wave equation transforms to the
standard Helmholtz equation. Also in the quantum scenario, for
stationary problems the unitary evolution of the wavefunction reduces
the Schr{\"o}dinger equation into the Helmholtz equation. Notable
example of the Dirichlet boundary condition (DBC) is the confinement
of a quantum particle in an infinite potential well where the
eigenfunction vanishes on the boundary. A nucleon in the nucleus can
be treated as this in the first approximation neglecting its
interaction with the other nucleons. Canonical example of Neumann
boundary condition (NBC) is the acoustic cavity where the sound
velocity (which is proportional to the gradient of the sound pressure)
is set to zero on the boundary. In order to find a solution of the
particular partial differential equation (PDE) we have to select, from
all the possible solutions of the aforesaid equation, the specific
combination which will respect the boundary condition. Now these
particular problems become different in the nature of the boundary
conditions imposed; either the geometry of the boundary varies or the
specific behaviour of the eigenfunction at the boundary is
different. So to categorise the solution of a particular PDE, say the
Helmholtz equation for our case, depending on the shapes of the
boundary surface and also according to the nature of the boundary
condition is an intricate task. For a restricted class of domains with
simple geometries in three dimensions, viz. rectangular, spherical and
cylindrical, analytic closed form solutions are available
\cite{p4ref13,p4ref12}. Constructing on the boundary a co-ordinate
system suitable to it often helps in finding the solution. For the
Helmholtz equation in three dimensions, there are $11$ such `special'
co-ordinate systems \cite{p4ref1}. However, even in these sacred
classes, the multi-parameter dependence of the solution makes them
difficult to use. So, for a general domain, finding solutions to the
Helmholtz equation becomes a formidable job. However, many physical
problems demand us to solve the equation for an arbitrary domain which
is significantly deviated from the above mentioned idealised
scenario. In particular, the geometry of the nanoscale second-phase
particles are described analytically as superellipsoids \cite{p5ref8}
and experimentally verified from just one micrograph \cite{p1ref1}. To
explore the shapes of these particles more accurately, we need a good
estimation of the eigenspectra of these geometries which urge us to
solve the Helmholtz equation for these arbitrary boundaries. Recently
in the field of medical science for automatic prostate segmentation
using deformable superellipses have been efficiently applied
\cite{p5ref9}. The analytic approach towards this problem came
recently by Panda and Hazra \cite{p5ref7} where they have generalised
Rayleigh's theorem in three dimensions and used the standard time
independent perturbation method of quantum mechanics to calculate the
eigenfunction and eigenvalue corrections in closed forms up to the
second order of perturbation for general shapes in three dimension in
terms of the spherical harmonics expansion coefficients. Most of the
earlier attempts made towards this problem were via numerical means in
the area of chaos for general three dimensional billiards
\cite{p4ref6,p4ref4,p4ref5,p4ref2,p4ref3,p4ref7} and also in some
recent experiments to study the wave chaos in microwave \cite{p4ref8}
and resonant optical cavities \cite{p4ref9}. Among the numerical
schemes, the finite difference method (FDM), the finite element method
(FEM) \cite{p5ref1} and the boundary element method (BEM)
\cite{p5ref2} are popular ones but they consume huge amount of time to
generate the mesh for a complicated geometry. Recently as alternatives
to the above, some meshless methods have been developed, like the
boundary collocation method \cite{p5ref6}, the radial basis function
\cite{p5ref5}, the method of particular solutions \cite{p5ref3} and
the method of fundamental solutions \cite{p5ref4}. These methods also
have the drawback of getting spurious eigensolutions. \\

In this paper, we have prescribed an alternative method to solve the
boundary value problem for the Helmholtz equation for a general simply
connected convex domain in three dimensions. This analytic formulation
is an extension of our earlier work for two dimensions \cite{p3ref4}
where a suitable diffeomorphism in terms of a Fourier series was
chosen to map the general problem into an equivalent one where the
boundary was circular but the equation gets complicated due to the
deformation of the metric in the interior and was solved by
perturbation technique of quantum mechanics.  In this method, an
arbitrary domain in three dimensions is mapped to a regular closed
region (in which the Helmholtz equation is exactly solvable) by a
suitable co-ordinate transformation resulting the deformation of the
interior metric. As a result, the homogeneous Helmholtz equation gets
modified and can be written as the original one with additional source
terms present on the right hand side of it. These extra terms can now
be treated as a perturbation to the original equation and is solved
using the standard perturbation technique. The corrections to the
eigenfunction are obtained in closed form at each order of
perturbation irrespective of the boundary conditions. The eigenvalue
correction at each order is calculated by applying the proper boundary
conditions on the respective eigenfunction corrections. To be able to
express the eigenfunction corrections independent of the boundary
conditions is a unique advantage of this method over the existing
ones. Also, in this method the boundary conditions maintain a simple
form at each order of perturbation and is easy to apply on some
regular closed surface. In this analysis, we have used spherical
harmonics to represent the asymmetries of the boundary from its
equivalent sphere which allows us to implement the method for a large
variety of geometries in three dimensions. To test the efficacy of the
perturbative method we have tabulated the analytical values with the
corresponding numerical ones for spheroids, supereggs, stadium of
revolution, rounded cylinder and pear shaped enclosures. The method
seems to work extremely well for these cases for both Dirichlet and
Neumann conditions. It has a potential of applicability for various
general shapes in three dimensions, where the deviations are large
from a spherical shape.\\ The paper is organised as follows : the
section \ref{form} describes the general formalism in the abstract
sense. In section \ref{app}, we have applied the method to various
enclosures. The last section \ref{res} is reserved for results,
conclusions and comments.

\section{Formulation}\label{form} 
The homogeneous Helmholtz equation reads,
\allowdisplaybreaks
\begin {equation}
\left(g_{ij} \nabla^{i} \nabla^{j} + k^{2}\right)\psi \equiv \left(\triangle + {\cal E}\right) \psi=0,   \label{eq:1}
\end {equation}
where $g_{ij}$ is the background metric component on three dimensional
space ($\cal V$) and $\nabla$ represents a covariant
derivative. $\triangle$ is the Laplacian operator in three
dimensions. Our interest lies in finding the eigenspectrum and the
solutions in the interior of a closed simply connected convex region
satisfying either the Dirichlet boundary condition (DBC), $\psi=0$, or
the Neumann boundary condition (NBC), $\frac{\partial \psi}{\partial n}
= 0$, on the boundary $\partial \cal V$, where $\frac{\partial
  \psi}{\partial n}$ is the derivative along the outward normal
direction to $\partial \cal V$. Depending on the boundary condition
the parameter $k^{2}$ can be matched to a physical quantity (for the
DBC it is proportional to the energy of a quantum particle confined
within $\partial \cal V$, where as for the NBC it will determine the
square of the resonating frequencies for acoustical cavities).

For convenience, we choose to work in spherical polar coordinate
system ($R, \Theta, \Phi$). Now, We consider a general arbitrary
surface of the form $R = R(\Theta, \Phi)$. The periodicity condition
implies that $R(\Theta , \Phi + 2\pi) = R(\Theta, \Phi)$. We assume
that any general shape (which is not very elongated in one direction)
can be expressed as a deformation around an effective spherical
boundary. We choose spherical boundary because the Helmholtz equation
in 3D is exactly solvable for this boundary (the analysis in principle
can work for perturbation around any simple surface for which the
Helmholtz equation is exactly solvable). We make a coordinate
transformation from the old one $(R, \Theta, \Phi)$ to a new one $(r,
\theta, \phi)$ of the form
\begin{subequations}
\label{eq:3}
\begin{align}
&R = r~(1 + \epsilon f(\theta,\phi))\\
&\Theta = \theta\\
&\Phi=\phi
\end{align}
\end{subequations}
where $\epsilon$ is a deformation parameter. Now, for an intelligent
choice of the well behaved function $f(\theta,\phi)$, our arbitrary
surface will be transformed into a sphere of average radius,
\begin{align}
R_{0} =\frac{1}{4\pi} \int \limits_{0}^{\pi} \int \limits_{0}^{2 \pi}
R(\theta,\phi) \sin \theta \text{d} \theta \text{d} \phi, \label{eq:avgrad} 
\end{align}
in the $(r,\theta,\phi)$ system. As a result of the deformation, there
will be changes in the underlying metric component $g_{ij}(R, \Theta,
\Phi)$. Henceforth, we use the notation
\begin{align*}
f^{(i,j)} &\equiv \frac{\partial^{i+j} f}{\partial \theta^{i} \partial
  \phi^{j}}.
\end{align*}
The dependence on the arguments $(\theta, \phi)$ is not shown
explicitly for brevity. The flat background metric in the old
coordinate system ($R, \Theta, \Phi$) is given by $g = {\rm diag}\,
(1, R^{2}, R^2 \sin^2 \Theta)$. Under the coordinate transformations
({\ref{eq:3}}) it takes the form
\begin{equation}
 \tilde{g} = \left(\begin{array}{lcc} (1+\epsilon f)^2 & r
   \epsilon (1+\epsilon f) f^{(1,0)} & r \epsilon (1+\epsilon f)
   f^{(0,1)} \\ r \epsilon (1+\epsilon f) f^{(1,0)} & r^2
   \left[(1+\epsilon f)^2+\epsilon ^2 f^{{(1,0)}^{2}}\right] & r^2
   \epsilon ^2 f^{(0,1)} f^{(1,0)}\\ r \epsilon (1+\epsilon f)
   f^{(0,1)} & r^2 \epsilon ^2 f^{(0,1)} f^{(1,0)} & r^2
   \left[(1+\epsilon f)^2 \sin^2 \theta +\epsilon ^2
     f^{{(0,1)}^{2}}\right]
\end{array}\right).\nonumber 
\end{equation}
The non-vanishing components of connection $\Gamma$ for the flat
background metric ($g$) given by,
\begin{align*}
\begin{array}{lll}
\Gamma^{R}_{\phantom r \Theta \Theta} = -R, &
\Gamma^{\Theta}_{\phantom r R \Theta} = \frac{1}{R}, &
\Gamma^{\Phi}_{\phantom r R \Phi} = \frac{1}{R},
\\ \Gamma^{R}_{\phantom r \Phi \Phi} = -R \sin^{2} \Theta, &
\Gamma^{\Theta}_{\phantom r \Phi \Phi} = -\cos \Theta \sin \Theta, &
\Gamma^{\Phi}_{\phantom r \Theta \Phi} = \cot \Theta,
\end{array}
\end{align*}
modify to
\begin{align*}
&\Gamma^{r}_{\phantom r \theta \theta} =~ -r\left[1 - \frac{\epsilon
      f^{(2,0)}}{(1+\epsilon f)} + \frac{2 \epsilon^2 f^{{(1,0)}^{2}}}{(1+\epsilon f)^2}\right],~  \Gamma^{r}_{\phantom r \theta \phi} =~ \epsilon r
  \left[\frac{\left(f^{(1,1)}-f^{(0,1)}\cot \theta\right)}{(1+\epsilon
      f)}- \frac{2 \epsilon f^{(1,0)}f^{(0,1)}}{(1+\epsilon f)^2}
    \right], \\ 
&\Gamma^{r}_{\phantom r \phi \phi} =~ -r\left[\sin^2
    \theta- \frac{ \epsilon \left(f^{(0,2)} + \cos \theta \sin \theta
      f^{(1,0)}\right)}{(1+\epsilon f)} +\frac{2 \epsilon^{2}
      f^{{(0,1)}^{2}}}{(1+\epsilon f)^2}  \right],~ \Gamma^{\theta}_{\phantom r r \theta} = \frac{1}{r},
  ~\Gamma^{\theta}_{\phantom r \theta \theta} = \frac{2 \epsilon
    f^{(1,0)}}{1+\epsilon f} \\
&\Gamma^{\theta}_{\phantom r \theta
    \phi} = \frac{\epsilon f^{(0,1)}}{1+\epsilon f},
  ~\Gamma^{\theta}_{\phantom r \phi \phi} = -\cos \theta \sin \theta,
   ~\Gamma^{\phi}_{\phantom r r \phi} = \frac{1}{r},
  ~\Gamma^{\phi}_{\phantom r \theta \phi} = \cot \theta
  +\frac{\epsilon f^{(1,0)}}{1+\epsilon f}, ~\Gamma^{\phi}_{\phantom r
    \phi \phi} = \frac{2 \epsilon f^{(0,1)}}{1+\epsilon f}.
\end{align*} 
As a result of the coordinate transformation ({\ref{eq:3}}) no
spurious curvature is induced in the manifold (i.e. Riemann tensor,
$R^{i}_{{\phantom i} jkl} = 0~~\forall~ i, j, k, l$). Under the map
$(R, \Theta, \Phi) \rightarrow (r, \theta, \phi)$, Eq. ({\ref{eq:1}})
takes the following form
\begin{flalign}
\label{eq:6}
&\frac{1}{r^2 (1+\epsilon f)^4}\left[\cot \theta (1+\epsilon f)^2
  \psi^{(0,1,0)}+r \left\{2(1+\epsilon f)^2 - \epsilon (1+\epsilon
  f)\left(\frac{f^{(0,2)}}{\sin^2 \theta} + \cot \theta f^{(1,0)}
  +f^{(2,0)}
  \right)\right.\right.\nonumber\\ &\left.\left.+2\epsilon^2\left(\frac{f^{{(0,1)}^{2}}}{\sin^2
    \theta}+f^{{(1,0)}^{2}}\right)\right\} \psi^{(1,0,0)}+r^2
  \left\{(1+\epsilon f)^2 + \epsilon^2 \left(
  \frac{f^{{(0,1)}^{2}}}{\sin^2 \theta} + f^{{(1,0)}^{2}}\right)
  \right\} \psi^{(2,0,0)} \right. \nonumber\\ &\left.+(1+\epsilon f)^2\left(\frac{\psi^{(0,0,2)}}{\sin^2 \theta} + \psi^{(0,2,0)}
  \right)- 2 r \epsilon(1+\epsilon f) \left(\frac{f^{(0,1)} }{\sin^2
    \theta}\psi^{(1,0,1)} + f^{(1,0)} \psi^{(1,1,0)} \right) \right]+
      {\cal E} \psi = 0,
\end{flalign}
where $$\psi^{(i,j,k)} \equiv \frac{\partial^{i+j+k} \psi}{\partial
  r^{i} \partial \theta^{j} \partial \phi^{k}}.$$ After some
simplification, Eq. ({\ref{eq:6}}) reduces to
\begin{equation}
\sum_{n = 0}^{\infty} \epsilon^{n} {\cal H}_{n}\psi  +   {\cal E} \psi = 0, \label{eq:7}
\end{equation}
where the operator ${\cal H}_{i}$'s are given by 
\begin{flalign}
{\cal H}_{0}\psi &=  {\cal D}^2\psi + \frac{1}{r^2}{\cal L}^2\psi\,, \\
{\cal H}_{1}\psi &= -\frac{1}{r} \Omega^{2}\psi - 2 f {\cal H}_{0}\psi\,,\\
{\cal H}_{2}\psi &=\frac{3f}{r} \Omega^{2}\psi +  {\cal F} {\cal D}^2\psi+ 3 f^2 {\cal H}_{0}\psi\,,\\
&\qquad\vdots \nonumber \\
{\cal H}_{m}\psi & = \frac{(-1)^m}{6} (m+1)f^{m-2} \left[ m(m-1)
  {\cal F} {\cal D}^2\psi+ 6 f^2 {\cal
    H}_{0}\psi + \frac{3mf}{r} \Omega^{2}\psi \right]   
\end{flalign}
and
\begin{flalign}
{\cal D}^2 \equiv &~\frac{\partial^2}{\partial r^2} + \frac{2}{r}
\frac{\partial}{\partial r}~;\quad
{\cal L}^2 \equiv ~\frac{\partial^2}{\partial \theta^2} + \cot \theta
\frac{\partial}{\partial \theta} + \frac{1}{\sin^2
  \theta}\frac{\partial^2}{\partial \phi^2}~, \nonumber\\  
\Omega^{2} \equiv &~\left[({\cal L}^2 f) + 2 f^{(1,0)}
  \frac{\partial }{\partial \theta} +
\frac{2 f^{(0,1)}}{\sin^2 \theta}\frac{\partial }{\partial \phi}
\right]\frac{\partial}{\partial r}~; \quad 
{\cal F} = ~f^{{(1,0)}^{2}} + \frac{f^{{(0,1)}^{2}}}{\sin^2 \theta}.\nonumber
\end{flalign}
We will now implement the standard Rayleigh-Schr{\"o}dinger
perturbation theory to solve for $\psi$ and ${\cal E}$. Perturbed
eigenfunction $\psi$ and the corresponding eigenvalue ${\cal E}$ are
expanded in a power series of the perturbation parameter, $\epsilon$,
as
\begin{subequations}
\begin{align}
\psi &= \psi^{(0)} + \epsilon \psi^{(1)} + \epsilon^{2} \psi^{(2)}
+\cdots ;  \\
{\cal E} &= {\cal E}^{(0)} + \epsilon {\cal E}^{(1)} + \epsilon^{2} {\cal E}^{(2)} +\cdots, 
\end{align}\label{eq:12}
\end{subequations}
where superscripts denote the orders of perturbation.

Substituting Eq. ({\ref{eq:12}}) in Eq. (\ref{eq:7}), and setting the
coefficients of different orders of $\epsilon$ to zero, yields
\begin{subequations}
\begin{align}
{\mathcal{O}}(\epsilon^{0})&: &&({\cal H}_{0} + {\cal E}^{(0)}) \psi^{(0)} = 0~,\label{eq:13} \\  
{\mathcal{O}}(\epsilon^{1})&:  &&({\cal H}_{0}+ {\cal E}^{(0)}) \psi^{(1)} + ({\cal H}_{1}
 + {\cal E}^{(1)}) \psi^{(0)} = 0~, \label{eq:14}\\ 
{\mathcal {O}}(\epsilon^{2})&: &&({\cal H}_{0} +{\cal E}^{(0)})  \psi^{(2)} +
({\cal H}_{1}+ {\cal E}^{(1)}) \psi^{(1)}+({\cal H}_{2} + {\cal E}^{(2)}) \psi^{(0)}  =
0~, \label{eq:15} \\
\vdots \nonumber\\
{\mathcal{O}}(\epsilon^{m})&:  &&\sum_{n = 0}^{m}\left({\cal H}_{n} + {\cal E}^{(n)}\right)
\psi^{(m-n)} = 0~.\label{eq:15a}
\end{align}
\label{eq:15tot}
\end{subequations}
We can infer from the above equations, \eqref{eq:15tot}, that new
source terms have been generated at each order in $\epsilon$ to the
unperturbed homogeneous Helmholtz equation as a by product of the
mapping given by ({\ref{eq:3}}). So, we started with an arbitrary
boundary with the absence of sources inside the domain and with a
mapping effectively generated a spherical boundary with non-vanishing
sources inside it. In order to maintain the simple boundary condition,
we have incorporated new source terms in the equations.\\ Now, the
eigenvalue corrections for different orders can be calculated by
\begin{subequations}
\begin{align}
{\cal E}^{(0)} = & -\langle \psi^{(0)}|{\cal H}_{0}|\psi^{(0)} \rangle ; \label{eq:16a} \\
{\cal E}^{(1)} = & -\langle \psi^{(0)}|{\cal H}_{1}|\psi^{(0)}\rangle; \label{eq:16b}   \\
{\cal E}^{(2)} = & -\langle \psi^{(0)}|{\cal H}_{1} + {\cal E}^{(1)} |\psi^{(1)} \rangle -
  \langle \psi^{(0)}|{\cal H}_{2}|\psi^{(0)} \rangle. \label{eq:16} \\
\vdots \nonumber\\
{\cal E}^{(m)} = & -\Big{\langle}
\psi^{(0)}\Bigl\lvert\sum_{n=1}^{m-1}\left({\cal H}_{n} +
{\cal E}^{(n)}\right)\Bigr\rvert \psi^{(m-n)} \Big{\rangle} -
  \Big{\langle} \psi^{(0)} \Bigl\lvert {\cal H}_{m} \Bigr\rvert \psi^{(0)} \Big{\rangle} . \label{eq:16c}
\end{align} 
\label{eq:16tot}
\end{subequations}
The corresponding boundary conditions for the DBC and the NBC are
respectively
\begin{align}
\psi^{(i)}(R_{0},\theta,\phi) =0\label{eq:dbctot}
\end{align}
and
\begin{align}
&\left[\frac{\partial\psi^{(i)}}{\partial r} +\left(f
  \frac{\partial}{\partial r} -\frac{f^{(1,0)}}{r}
  \frac{\partial}{\partial \theta}- \frac{f^{(0,1)}}{r
    \sin^{2} \theta} \frac{\partial}{\partial
    \phi}\right)\psi^{(i-1)}+ {\cal F} \sum
  \limits_{n=0}^{i-3}(-1)^{n}f^{n}\frac{\partial\psi^{(i-n-2)}}{\partial
    r}\right]\Bigg\rvert_{(R_{0},\theta,\phi)} = 0.\label{eq:nbctot}
\end{align}
%\begin{align}
%&\left(\frac{\partial\psi^{(i)}}{\partial r} + f
%  \frac{\partial\psi^{(i-1)}}{\partial r} -\frac{f^{(1,0)}}{r}
%  \frac{\partial\psi^{(i-1)}}{\partial \theta}- \frac{f^{(0,1)}}{r
%    \sin^{2} \theta} \frac{\partial\psi^{(i-1)}}{\partial
%    \phi}\right.\nonumber\\ &\left.\qquad\quad~+\left\{f^{{(1,0)}^2}+\frac{f^{{(0,1)}^2}}{\sin^{2}
%    \theta}\right\}\sum
%  \limits_{n=0}^{i-3}(-1)^{n}f^{n}\frac{\partial\psi^{(i-n-2)}}{\partial
%    r}\right)\Bigg\rvert_{(R_{0},\theta,\phi)} = 0.\label{eq:nbctot}
%\end{align}
for all $i \in \mathbb{N}$, where the radius of the sphere $R_{0}$ is
defined by \eqref{eq:avgrad}.\\ The general solution of the
Eq. ({\ref{eq:13}}) is given by
\begin{subequations}
\begin{align}
\psi_{n,l,m}^{(0)}  & =  N_{n,0}~ j_{0}(\rho)\,,  \qquad\qquad \qquad \qquad ~~~(l = 0)\,; \label{eq:21} \\
         & =  N_{n,l}~ j_{l}(\rho) Y^{m}_{l} , \qquad \qquad\qquad\qquad (l \neq 0)\,, 
         \label{eq:21a}
\end{align}
\end{subequations}
where $N_{n,l}$ is a suitable normalisation constant with $l \in
\mathbb{N}$, $n \in \mathbb{N}^{+}$ and $m =\{-l,-(l-1), \cdots, 0,
\cdots,l-1, l\} \in \mathbb{Z}$. $j_{l}(\rho)$ is the $l^{\text{th}}$
order spherical Bessel function of the first kind with the argument
$\rho = r \sqrt{{\cal E}_{n,l}^{(0)}} $, where ${\cal E}_{n,l}^{(0)}$
are the eigenvalues of the unperturbed Helmholtz equation. $Y^{m}_{l}$
is the spherical harmonics of order $l$ and degree $m$. The
expressions for the normalisation constant and the unperturbed
eigenvalues will be distinct for different boundary conditions but the
form of the general solution remains the same. Henceforth, we will
discuss both the cases, viz. the Dirichlet and the Neumann boundary
condition parallely. For the DBC, the eigenvalue ${\cal
  E}_{n,l}^{(0)}$ is calculated using the $ n^{\text{th}}$
zero\footnote{Here we have used an unusual notation for zeros of
  spherical Bessel functions and their derivatives. Conventionally,
  $a_{_{l,n}}$ and $a^{\prime}_{_{l,n}}$ denote $n^{\text{th}}$ zeros
  of $j_{l}$ and $j^{\prime}_{l}$ respectively.} of $j_{l}$, denoted
by $\beta_{_{n,l}}$, and for the NBC, it is dictated by the $
n^{\text{th}}$ zero of $j^{\prime}_{l}$ (i.e. derivative of $j_{l}$
with respect to its argument), denoted by $\alpha_{_{n,l}}$. The
boundary conditions Eq. ({\ref{eq:dbctot}}) and
Eq. ({\ref{eq:nbctot}}) for $i = 0$ imply
\begin{align}
 {\cal E}_{n,l}^{(0)} =~&\beta^{2}_{_{n,l}}/R_{0}^{2} \,,~\qquad \qquad \mathrm{(DBC)} ~;\label{eq:endbc}\\
             =~&\alpha^{2}_{_{n,l}}/R_{0}^{2}\,,\qquad \qquad ~\mathrm{(NBC)}~,\label{eq:ennbc}
\end{align}
where all the levels with non-zero $l$ are $(2l+1)$-fold degenerate.

In this prescription, the energy corrections can be obtained in two
ways. Primarily, ${\cal E}^{(i)}$ is extracted out by imposing the
respective boundary conditions on $\psi^{(i)}$ given by
Eqs. (\ref{eq:dbctot}) and Eqs. (\ref{eq:nbctot}), which as a bonus
give the coefficients of Bessel functions (in
$\psi^{(i)}$). Alternatively, it can be verified using
Eqs. (\ref{eq:16tot}) from the information of $\psi^{(m)}(\forall~ m <
i)$. In principle, this formulation can be applied to obtain
correction at all order of perturbation. In the following we calculate
the eigenvalue as well as eigenfunction corrections for both the cases
of boundary conditions. Till now the formalism was proceeding in an
abstract sense as it did not require a particular form of $f$. From
now on without a loss of generality we choose the following
specific form for $f$ in terms of spherical harmonics
\begin{align}
f = \sum \limits_{a=1}^{\infty} \sum \limits_{b=-a}^{a} C^{b}_{a}
Y^{b}_{a}, \label{eq:fexp}
\end{align}
where $C^{b}_{a}$ are the expansion coefficients. The constant part
$C^{0}_{0}$ can always be absorbed by redefining the $R$ in
\eqref{eq:3}.  \allowdisplaybreaks
\subsection{Non-degenerate states ($l=0$)} 
The first order correction to the eigenfunction is obtained by solving
the Eq. ({\ref{eq:14}}). Thus, we have
\begin{align}
\psi_{n,0,0}^{(1)} =&\, A_{0} j_{0}(\rho) -\frac{N_{n,0}}{2} E_{n,0}^{(1)}\rho j_{1}(\rho) +
\sum_{p=1}^{\infty} \sum_{q=-p}^{+p} A^{q}_{p} j_{p}(\rho) Y^{q}_{p} -
N_{n,0} f \rho j_{1}(\rho),   \label{eq:22} 
\end{align}
where the unknown coefficients $ E_{n,0}^{(1)}\,(={\cal
  E}_{n,0}^{(1)}/{\cal E}_{n,0}^{(0)})$,
$A_{0}\,(=A^{0}_{0}Y^{0}_{0})$ and $A^{q}_{p}$ will be calculated by
imposing the respective boundary conditions. The terms containing
$N_{n,0} \rho j_{1}(\rho)$ make the particular integral of the
Eq. ({\ref{eq:14}}). Now, applying the boundary conditions, given by
Eq. ({\ref{eq:dbctot}}) and Eq. ({\ref{eq:nbctot}}), for $i=1$, we
extract out the first order eigenvalue corrections as well as the
unknown expansion coefficients as
\begin{align*}
&{\cal E}_{n,0}^{(1)} = 0 ~~~\mbox{(for both the cases)};
  \\ 
&A^{q}_{p} = \beta_{_{n,0}} N_{n,0}
  C^{q}_{p}j_{1}(\beta_{_{n,0}})/j_{p}(\beta_{_{n,0}})\,, ~ (p \neq
  0)\qquad \mbox{(DBC)};\\
&A^{q}_{p} = \alpha_{_{n,0}} N_{n,0} C^{q}_{p}
  j_{0}(\alpha_{_{n,0}})/j^{\prime}_{p}(\alpha_{_{n,0}})\,, ~ (p \neq
  0)\qquad \mbox{(NBC)}.
\end{align*}
It is evident that the first order eigenvalue correction is zero for
both the cases of boundary conditions which is in confirmation with
Eq. \eqref{eq:16b}. The orthogonality relation between
$\psi_{n,0,0}^{(0)}$ and $\psi_{n,0,0}^{(1)}$ dictates that the
remaining constant $A_{0}$ of Eq. ({\ref{eq:22}}) is zero for both the
boundary conditions.

The eigenfunction correction for the second order is given by solving
Eq. \eqref{eq:15} as
\begin{align}
&\psi_{n,0,0}^{(2)} = B_{0} j_{0}(\rho)-\left(E_{n,0}^{(1)}A_{0}+E_{n,0}^{(2)}N_{n,0}+2A_{0}f\right)\frac{\rho
    j_{1}(\rho)}{2} + N_{n,0} E_{n,0}^{{(1)}^{2}} \frac{\rho^{2}j_{2}(\rho)}{8}-\frac{N_{n,0}}{2}
  \rho^2 j^{\prime}_{1}(\rho) f^2 \nonumber\\
&-\frac{N_{n,0}}{2} E_{n,0}^{(1)} \left\{\rho
    j_{1}(\rho)+\rho^2 j^{\prime}_{1}(\rho)\right\}f +\sum \limits_{p=1}^{\infty}\sum
  \limits_{q=-p}^{p}\left[B^{q}_{p} j_{p}(\rho) + A^{q}_{p} \left(f+\frac{1}{2}E_{n,0}^{(1)}\right)\rho
  j^{\prime}_{p}(\rho)\right]Y^{q}_{p}.
\label{eq:24}
\end{align}
The first non-zero eigenvalue correction ${\cal E}_{n,0}^{(2)}$ is
obtained with the knowledge of $\psi_{n,0,0}^{(0)}$ and
$\psi_{n,0,0}^{(1)}$ and imposing the boundary conditions,
Eq. (\ref{eq:dbctot}) and Eq. (\ref{eq:nbctot}), for $i = 2$, one
extracts out
\begin{align}
E_{n,0}^{(2)} =\frac{{\cal E}_{n,0}^{(2)}}{{\cal
    E}_{n,0}^{(0)}} = \sum \limits_{p=1}^{\infty}\sum
\limits_{q=-p}^{p} \frac{(-1)^{q}}{2\pi} C_{p}^{q} C_{p}^{-q}  \xi_{n,p}   \,; ~~
\xi_{n,p}  = 1 +  \frac{\beta_{_{n,0}}
  j^{\prime}_{p}(\beta_{_{n,0}})}{j_{p}(\beta_{_{n,0}})},\label{eq:27} ~~~\text{(DBC)}
\end{align}
and
\begin{align}
E_{n,0}^{(2)} =\frac{{\cal E}_{n,0}^{(2)}}{{\cal E}_{n,0}^{(0)}}= -\, \sum
\limits_{p=1}^{\infty}\sum \limits_{q=-p}^{p} \frac{(-1)^{q}}{2\pi}
C_{p}^{q} C_{p}^{-q} \lambda_{n,p} \,; ~ \lambda_{n,p} = 1 +
\frac{\alpha_{_{n,0}}
  j_{p}(\alpha_{_{n,0}})}{j^{\prime}_{p}(\alpha_{_{n,0}})}.\label{eq:27nb} ~~~~\text{(NBC)}
\end{align}
The coefficients $B^{q}_{p}$ are determined as
\begin{align*}
& B^{q}_{p} =  \frac{\beta_{_{n,0}}
  j_1(\beta_{_{n,0}})}{j_p(\beta_{_{n,0}})} A_{0}C_{p}^{q} -
\frac{N_{n,0} \beta_{_{n,0}} j_1(\beta_{_{n,0}})}{j_p(\beta_{_{n,0}})}
\sum_{a=1}^{\infty} \sum_{b=-a}^{+a} \sum^{a+p}\limits_{k=\left \lceil
  \substack{|a-p|\\|q-b|}\right \rceil }
\sqrt{\frac{(2a+1)(2k+1)}{4\pi(2p+1)}}~ C_{a}^{b} C_{k}^{q-b}
\nonumber\\ 
&\times \langle a k 0 0|p 0 \rangle \langle a k b (q-b)|p
q \rangle~ \xi_{n,a}\,, \qquad\qquad\qquad\qquad\qquad\qquad\qquad\qquad\qquad\qquad\quad\text{(DBC)} \\ 
&= \frac{\alpha_{_{n,0}}
  j_0(\alpha_{_{n,0}})}{j^{\prime}_p(\alpha_{_{n,0}})} A_{0}C_{p}^{q}
+ \frac{N_{n,0} \alpha_{_{n,0}}
  j_0(\alpha_{_{n,0}})}{j^{\prime}_p(\alpha_{_{n,0}})}
\sum_{a=1}^{\infty} \sum_{b=-a}^{+a} \sum^{a+p}\limits_{k=\left \lceil
  \substack{|a-p|\\|q-b|}\right \rceil }
\sqrt{\frac{(2a+1)(2k+1)}{4\pi(2p+1)}}~ C_{a}^{b} C_{k}^{q-b} 
\nonumber\\ 
& \times \langle a k 0 0|p 0 \rangle \langle a k b (q-b)|p q \rangle \left\{\lambda_{n,a} + \frac{k(k+1)-a(a+1)-p(p+1)}{2 \alpha_{_{n,0}}}
\frac{j_a(\alpha_{_{n,0}})}{j^{\prime}_a(\alpha_{_{n,0}})}\right\},\quad \text{(NBC)}
\end{align*}
where $\langle n_{1} n_{2} m_{1} m_{2}|n m \rangle $ gives the
Clebsch-Gordan coefficient for the decomposition of $| n_{1}, n_{2},
n, m \rangle$ in terms of $|n_{1}, m_{1}\rangle$ $|n_{2},
m_{2}\rangle$ and $\left \lceil \substack{|a-p|\\|q-b|}\right \rceil
\equiv$ Maximum ($|a-p|,|q-b|$). The unknown coefficient $B_{0}$ can
be fixed by the normalisation condition of the eigenfunction corrected
up to second order. These results are consistent with the earlier work
done by the other method \cite{p5ref7}. The similarities of these
results with the two dimensional metric deformation results
\cite{p3ref4} can also be noticed.
\subsection{Degenerate states ($l \neq 0$)}
In order to bypass the complexity, we consider only the problem of
axisymmetric domains for the degenerate states, $l \neq 0$. This
should not be a major handicap for the formalism as the shapes
concerned in variety of physical problems are often axisymmetric
deviations from a sphere. Of course an $f$ with full spherical
harmonics can be dealt in the same fashion. With this simplification
the expansion for $f$ given in \eqref{eq:fexp} will reduce to
\begin{align}
f = \sum \limits_{a=1}^{\infty} C_{a} Y^{0}_{a}(\theta),
\end{align}
where we denote $C_{a}\equiv C^{0}_{a}$. The above expression implies
that the boundary has azimuthal symmetry. Due to this simplification
the boundary conditions for NBC in \eqref{eq:nbctot} will reduce
further and all the terms containing $\phi$-derivative of $f$ will go
to zero for the axisymmetric geometries.  For $l \neq 0$ case, the
first order eigenfunction correction, given by
\begin{align}
\psi_{n,l,m}^{(1)} =&\,
\sum_{p=0}^{\infty} \sum_{q=-p}^{+p} A^{q}_{p} j_{p}(\rho) Y^{q}_{p}+
N_{n,l} \rho j^{\prime}_{l}(\rho) f Y^{m}_{l} +
\frac{N_{n,l}}{2} E_{n,l}^{(1)} \rho j^{\prime}_{l}(\rho) Y^{m}_{l}, \label{eq:23}
\end{align}
is a solution of Eq. ({\ref{eq:14}}) with $\psi_{n,l,m}^{(0)}$ given
by \eqref{eq:21a}. The first order eigenvalue correction is evaluated
by setting the respective boundary conditions given in
Eq. (\ref{eq:dbctot}) and Eq. (\ref{eq:nbctot}) for $i =1$ and it is
also confirmed with the Eq. ({\ref{eq:16b}}). Thus, we have
\begin{align}
E_{n,l}^{(1)}=& \frac{{\cal E}_{n,l}^{(1)}}{{\cal E}_{n,l}^{(0)}} =
-\sum \limits_{k=1}^{l} \sqrt{\frac{(4k+1)}{\pi}}~ C_{2k}~ \langle
(2k) l 0 0| l 0 \rangle \langle (2k) l 0 m| l m \rangle
,\qquad\qquad\qquad\qquad\quad \mbox{(DBC)}~;\nonumber \\ = &-\sum
\limits_{k=1}^{l} \sqrt{\frac{(4k+1)}{\pi}}~ C_{2k}~ \langle (2k)
l 0 0| l 0 \rangle \langle (2k) l 0 m| l m \rangle \left( 1 +
\frac{k(2k+1)}{\alpha^{2}_{_{n,l}}
  -l(l+1)}\right),\qquad\mbox{(NBC)}~,\nonumber
\end{align}
where the corresponding unperturbed eigenvalues ${\cal E}_{n,l}^{(0)}$
are given by Eq. ({\ref{eq:endbc}}) and Eq. ({\ref{eq:ennbc}})
respectively. There is a non-zero correction to the eigenvalue even at
the first order unlike the non-degenerate case. Further, the
coefficients $A^{q}_{p}$s (for $p \neq l$) are extracted as,
\begin{align*}
A_{p}^{m} = &\frac{N_{n,l}\beta_{_{n,l}}
  j_{l+1}(\beta_{_{n,l}})}{j_p(\beta_{_{n,l}})} \sum \limits_{k=|l-p|}^{l+p}
\sqrt{\frac{(2k+1)(2l+1)}{4\pi (2p+1)}}~ C_{k} \langle k l 0 0|
p 0 \rangle \langle k l 0 m| p m \rangle, \qquad\qquad\quad
~\mathrm{(DBC)}\\
 =&\frac{N_{n,l}\alpha_{_{n,l}}
  j_{l}(\alpha_{_{n,l}})}{j^{\prime}_p(\alpha_{_{n,l}})} \sum
\limits_{k=|l-p|}^{l+p}
\sqrt{\frac{(2k+1)(2l+1)}{4\pi (2p+1)}}~ C_{k} \langle k l 0 0|
p 0 \rangle \langle k l 0 m| p m \rangle\\ &\qquad\qquad\times
\left\{\alpha^{2}_{_{n,l}} +\frac{k(k+1)-l(l+1)-p(p+1)}{2}\right
\},\qquad\qquad\qquad\qquad\qquad\quad\mathrm{(NBC)}
\end{align*}
where rest of the coefficients $A^{q}_{p}$ for $q \neq m$ are
zero. The remaining coefficient $A^{m}_{l}$ is calculated from the
normalisation condition and found to be
\begin{align*}
A^{m}_{l} = &~ 0,  \qquad \qquad \qquad \qquad\qquad \qquad \qquad \qquad \qquad\qquad\qquad \qquad \qquad \qquad \qquad~\text{(DBC)}\\
= &
-\frac{1}{8}\frac{\left[\alpha^{2}_{_{n,l}}-3l(l+1)\right]}{\left[\alpha^{2}_{_{n,l}}-l(l+1)\right]^{2}}\sum
\limits_{k=1}^{2l} \sqrt{\frac{(2k+1)}{\pi}}~ k(k+1)C_{k} \langle k l 0 0| l
0 \rangle \langle k l 0 m| l m \rangle. \qquad \text{(NBC)} 
\end{align*}
By solving Eq. ({\ref{eq:15}}) we get the second order correction to
the eigenfunction as
\begin{align*}
&\psi_{n,l,m}^{(2)} =\sum \limits_{p=0}^{\infty}\sum
  \limits_{q=-p}^{p}\left[B^{q}_{p} j_{p}(\rho) + A^{q}_{p}
    \left(f+\frac{1}{2}E_{n,l}^{(1)}\right)\rho
    j^{\prime}_{p}(\rho)\right]Y^{q}_{p}-\frac{E_{n,l}^{(2)}N_{n,l}}{2}\rho
  j_{l+1}(\rho)Y^{m}_{l} \\
&+\frac{N_{n,l}E_{n,l}^{(1)}}{2}\left[\frac{E_{n,l}^{(1)}}{4}\left\{\rho^{2}j_{l+2}(\rho)-2l\rho
    j_{l+1}(\rho)\right\}+\left\{\rho^{2}j_{l+2}(\rho)-2(l+1)\rho
    j_{l+1}(\rho)+l^{2}j_{l}(\rho) \right\}f\right]Y^{m}_{l}\\
&+\frac{N_{n,l}}{2}\left\{\rho^{2}j_{l+2}(\rho)-(2l+1)\rho
    j_{l+1}(\rho)+l(l-1)j_{l}(\rho)\right\}f^{2}Y^{m}_{l},
\end{align*}
where $E_{n,l}^{(2)}= {\cal E}_{n,l}^{(2)}/{\cal E}_{n,l}^{(0)}$.
Imposing the boundary conditions (\ref{eq:dbctot}) and
(\ref{eq:nbctot}), for $i = 2$, we obtain the second order correction
to the eigenvalue for the DBC and the NBC respectively as follows
\begin{align}
&E_{n,l}^{(2)} = \frac{E_{n,l}^{(1)^{2}}}{4} + \sum
  \limits_{a,s=1}^{\infty} \sum \limits_{k=|a-s|}^{2l}\frac{\sqrt{(2a+1)(2s+1)}}{2\pi}
  C_{a}C_{s} \langle a s 0 0| k 0 \rangle^{2} \langle k l 0 0| l 0
  \rangle \langle k l 0 m| l m \rangle \nonumber\\
 & + \sum \limits_{\substack{p=|m|\\p \neq l}}^{\infty} \sum
  \limits_{s,k=|l-p|}^{l+p} \frac{\sqrt{(2s+1)(2k+1)}}{2\pi}
  \frac{\beta_{_{n,l}}
    j^{\prime}_p(\beta_{_{n,l}})}{j_p(\beta_{_{n,l}})}C_{s}C_{k}
  \langle k l 0 0| p 0\rangle \langle k l 0 m|
  p m \rangle \langle s p 0 0| l 0 \rangle \langle s p 0 m| l m
  \rangle,\nonumber\\ %~\text{(DBC)}
 &E_{n,l}^{(2)}
  =\left(\frac{\alpha^{2}_{_{n,l}}-3l(l+1)}{\alpha^{2}_{_{n,l}}-l(l+1)}
  \right)\frac{E_{n,l}^{(1)^{2}}}{4}
  -\left(\frac{l(l+1)}{\alpha^{2}_{_{n,l}}-l(l+1)}\right)E_{n,l}^{(1)}
  \sum \limits_{k=1}^{l}\sqrt{\frac{(4k+1)}{\pi}}C_{2k}\langle
  (2k) l 0 0| l 0 \rangle\nonumber \\
 &\times \langle (2k) l 0 m| l m \rangle + \sum
  \limits_{a,s=1}^{\infty}\sum
  \limits_{k=|a-s|}^{2l}\frac{\sqrt{(2a+1)(2s+1)}}{2\pi} C_{a}C_{s}
  \langle a s 0 0| k 0 \rangle^{2} \langle k l 0 0| l 0 \rangle
  \langle k l 0 m| l m \rangle \nonumber\\
&\times \left(1+ \frac{k(k+1)-2l(l+1)}{2\{\alpha^{2}_{_{n,l}}- l(l+1)\}} \right)-\sum
  \limits_{\substack{p=|m|\\p \neq l}}^{\infty} \sum \limits_{s,k=|l-p|}^{l+p}
  \frac{\sqrt{(2s+1)(2k+1)}}{\pi} C_{s}C_{k} \langle k l 0 0| p
  0\rangle \langle k l 0 m| p m \rangle\nonumber\\
 &\times \langle s p 0 0| l 0 \rangle \langle s p 0 m| l m \rangle \left(1 +
  \frac{k(k+1)+l(l+1)-p(p+1)}{2\{\alpha^{2}_{_{n,l}}-l(l+1)\}}\right) \times
  \nonumber\\
 &\times 
  \left(1 +\frac{2\alpha^{2}_{_{n,l}}
    +s(s+1)-p(p+1)-l(l+1)}{4}\frac{j_p(\alpha_{_{n,l}})}{\alpha_{_{n,l}}
    j^{\prime}_p(\alpha_{_{n,l}})}\right).\nonumber %\qquad\qquad\qquad \text{(NBC)} 
\end{align}
The expansion coefficients $B^{q}_{p}$s are algebraically complicated
to calculate and are not needed for our present purpose. These results
are matching with the results obtained in our earlier paper
\cite{p5ref7} by a different method.
\section{Examples} \label{app}
In the previous section we have described the formalism in an abstract
sense and now we apply it to estimate the energyspectra for various
axisymmetric boundary surfaces like spheroidal, superegg, stadium of
revolution, rounded cylinder and pear shaped enclosures. These
geometries are naturally encountered in nuclear physics
\cite{nature13} and in the experiments on nanoscale structures
\cite{p1ref1}. The analytic results have been compared against the
numerical ones obtained by using finite element method (with the help
of Matlab and Mathematica) and are tabulated below for the above
mentioned domains satisfying different boundary conditions.
\subsection{Supereggs}
\begin{figure}[htb]
\centering
\subfloat[~$n=1.7$]{\includegraphics[width=0.25\textwidth,trim=0in 0.9in 0in 0.75in,clip=true]{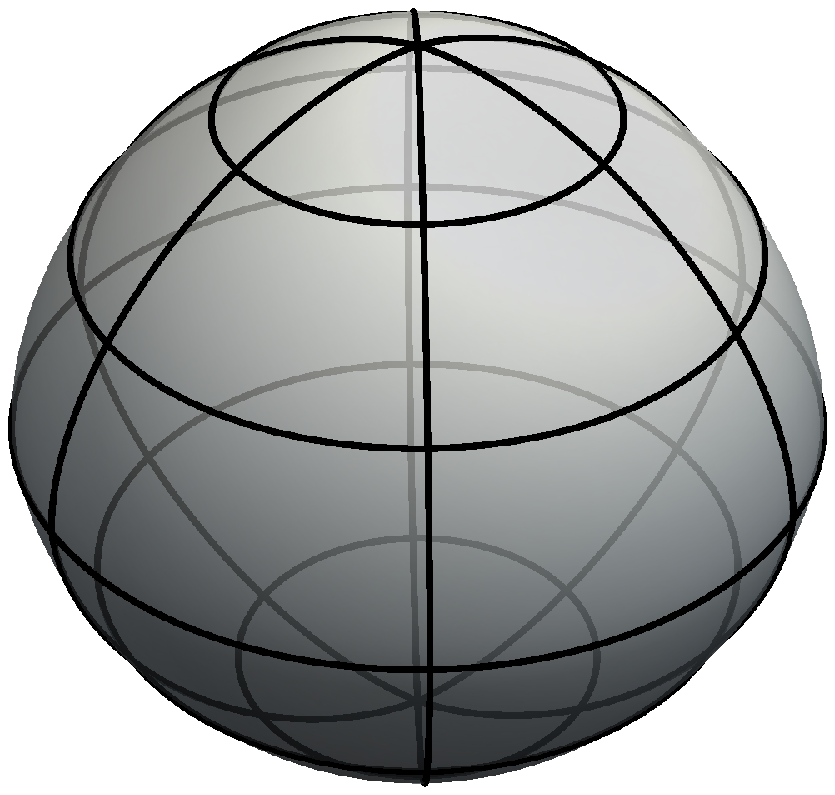}\label{fig:seg1}}
\subfloat[~$n=2.5$]{\includegraphics[width=0.25\textwidth,trim=0in 0.9in 0in 0.75in,clip=true]{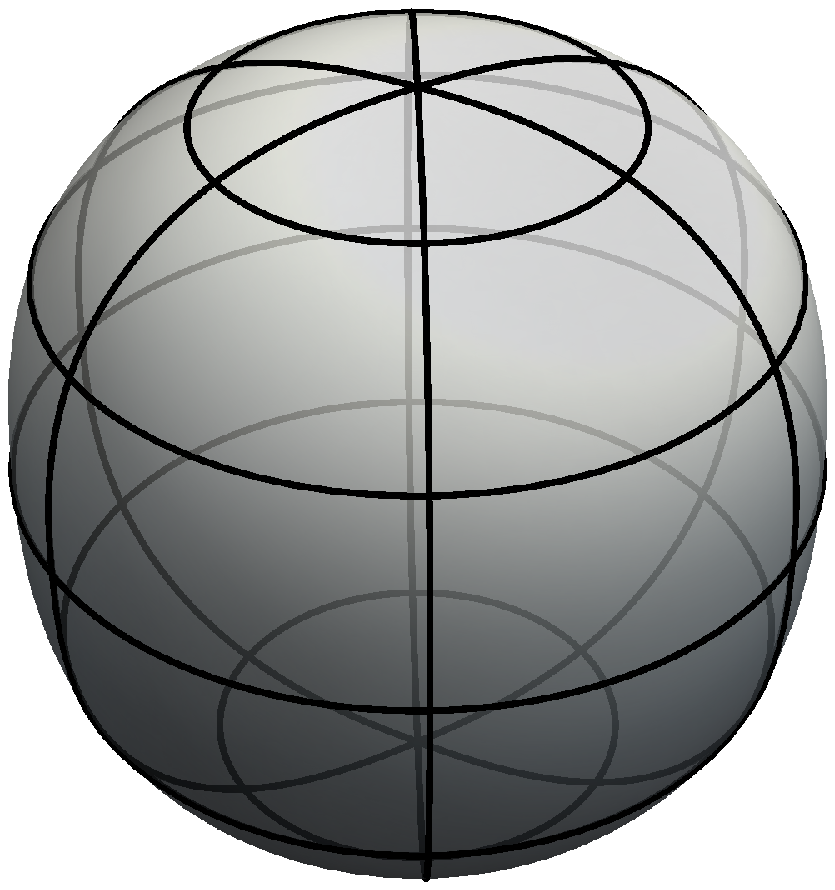}\label{fig:seg2}}
\caption{Supereggs for different exponents $n$.}
\label{fig:seg}
\end{figure}

We consider superegg \cite{p4ref11} shaped enclosures which
are surface of revolution of supercircles \cite{p1ref41} about
either of its in-plane axes. The representation of it in the spherical
polar coordinate is given by
\begin{equation}
r(\theta, \phi) = \frac{1}{\left( |\cos \theta|^{n} + |\sin \theta|^{n}\right)^{1/n}}, \label{eq:rseg}
\end{equation}
with the exponent $n>0$ and $\theta \in [0,\pi]$, $\phi \in [0,
  2\pi]$. Figs \eqref{fig:seg1} and \eqref{fig:seg2} depict the shapes
of the supereggs for $n=1.7$ and $2.5$ respectively. We have chosen
these two values, which lie on the opposite sides of the sphere (for
which $n=2$), to show the validity of the scheme on both extents.
\subsection{Rounded cylinder and Stadium of revolution}
\begin{figure}[htb]
\centering
\subfloat[~Rounded cylinder]{\includegraphics[width=0.25\textwidth,trim=0in
    0.5in 0in
    0.25in,clip=true]{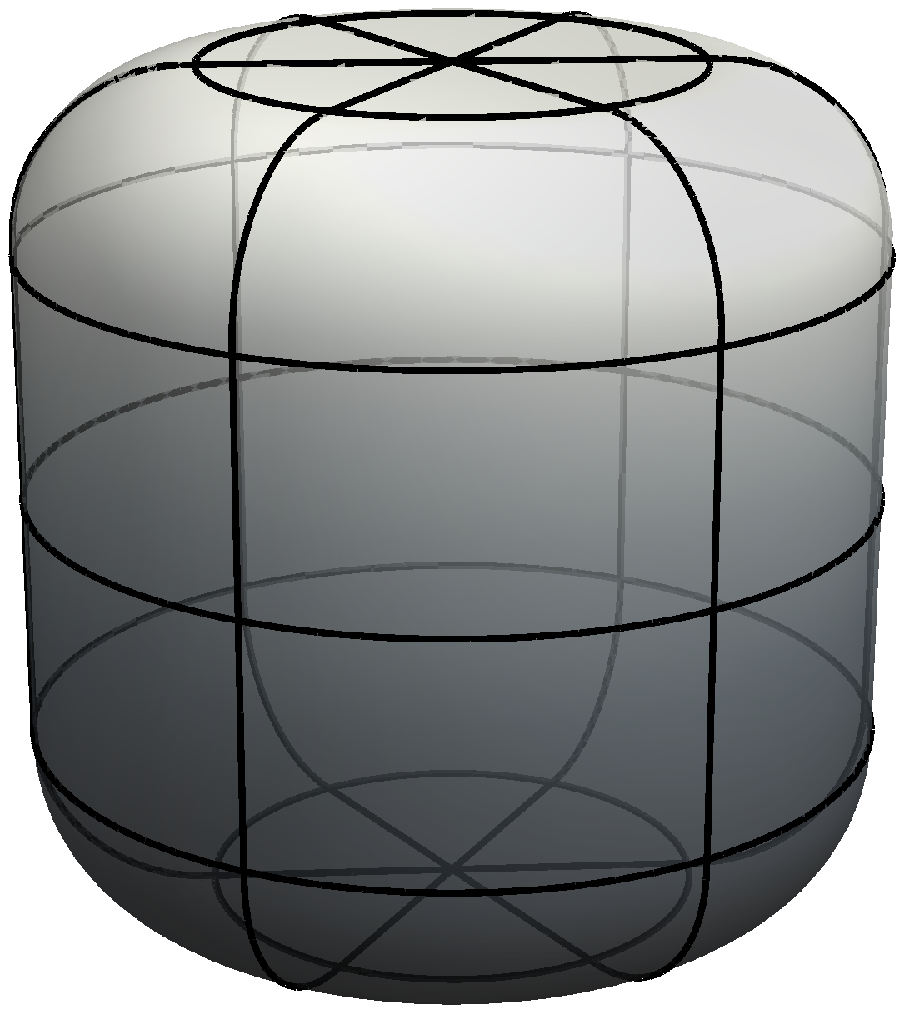}\label{fig:rcyld3}}
\subfloat[~Stadium of revolution]{\includegraphics[width=0.25\textwidth,trim=0in 0.75in 0in 0.5in,clip=true]{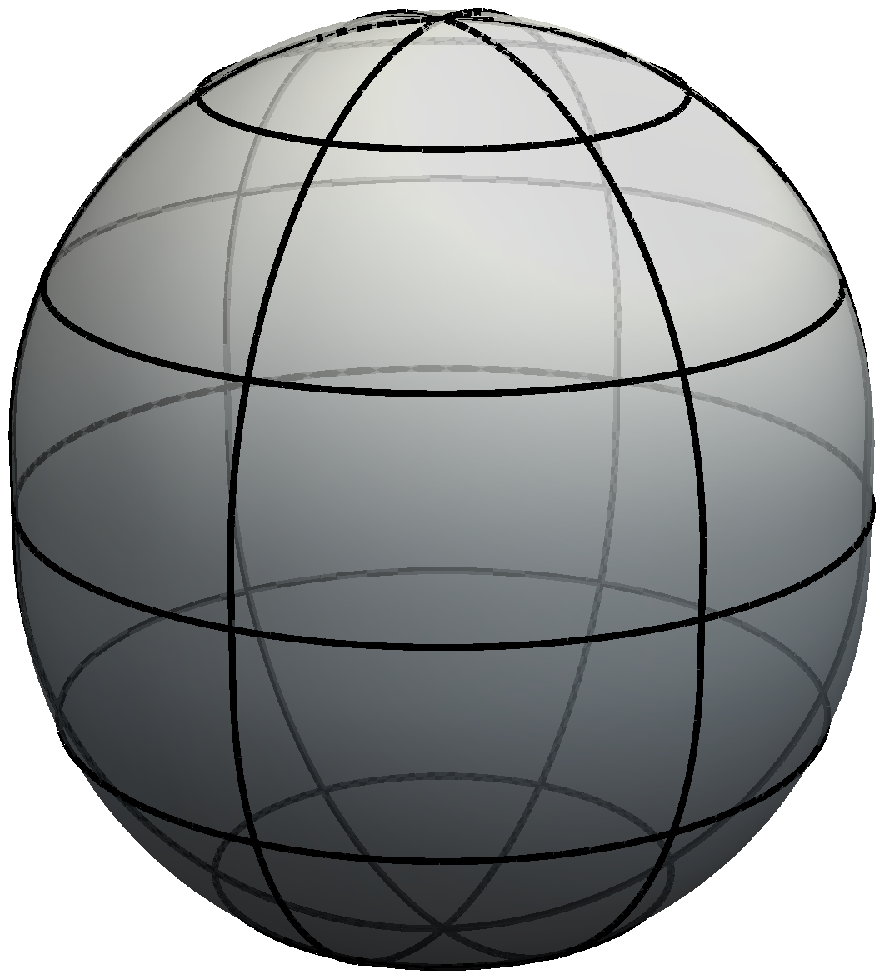}\label{fig:std3}}
\caption{The figure in the left \eqref{fig:rcyld3} is a rounded
  cylinder with the parameters $d=3\sqrt{3}/10$ and $R=2\sqrt{3}/10$
  while the right one \eqref{fig:std3} is the stadium of revolution
  with the parameters $d=1/4$ and $R=1$.}
\label{fig:stad}
\end{figure}

Next, we consider a rounded cylindrical enclosure in three
dimensions. If we consider a rounded rectangular region with circular
arc of radius $R$ and the separation in between them of $d$ in a
plane, then the surface of revolution around the major axis will be a
rounded cylinder of height and base diameter equal to $2(d+R)$. The
equation of the rounded rectangle in polar coordinate is
\begin{align}
{\mathcal R}(\theta)= \begin{cases}-\frac{d+R}{\sin \theta} & -\frac{\pi}{2} \le \theta
\le -\tan^{-1} (\frac{d+R}{d})\\
-d(\sin \theta -\cos \theta) + \sqrt{\left(R^2-d^2\right)-2 d^2 \sin 2\theta} & -\tan^{-1} (\frac{d+R}{d}) \le \theta \le -\tan^{-1} (\frac{d}{d+R})\\
\frac{d+R}{\cos \theta} & -\tan^{-1} (\frac{d}{d+R}) \le \theta
\le \tan^{-1} (\frac{d}{d+R}) \\
d(\sin \theta +\cos \theta) + \sqrt{\left(R^2-d^2\right)+ 2 d^2 \sin 2\theta} & 
\tan^{-1} (\frac{d}{d+R}) \le \theta \le \tan^{-1} (\frac{d+R}{d})\\
\frac{d+R}{\sin \theta} & \tan^{-1} (\frac{d+R}{d}) \le
\theta \le \frac{\pi}{2}
\end{cases}  \label{eq:rec}
\end{align}
 and the form of the rounded cylinder in spherical polar
 coordinate is
\begin{align}
r(\theta, \phi)= {\mathcal R}\left(\theta- \frac{\pi}{2}\right); \qquad ~\text{for}~~\theta \in
[0,\pi]~~\text{and}~~\phi \in [0, 2\pi].\label{eq:rcyld}
\end{align}
The shape of the rounded cylinder is displayed in fig
\eqref{fig:rcyld3}.  Further, we consider the stadium of revolution in
three dimensions. It is a surface of revolution attained by rotating a
half stadium shape in two dimensions about its major axis. The
equation of the half stadium in polar coordinate is
\begin{align}
{\mathcal S}(\theta)= \left\{\begin{array}{ll} -\frac{d}{2} \sin \theta +
\sqrt{R^{2}-\left(\frac{d}{2}\cos \theta \right)^{2}} & \qquad -\frac{\pi}{2} \le \theta
\le -\tan^{-1} (\frac{d}{2R})\\
\frac{R}{\cos \theta} & \qquad -\tan^{-1} (\frac{d}{2R}) \le \theta \le \tan^{-1} (\frac{d}{2R}) \\
\frac{d}{2} \sin \theta + \sqrt{R^{2}-\left(\frac{d}{2}\cos \theta
  \right)^{2}} & \qquad  \tan^{-1} (\frac{d}{2R}) \le
\theta \le \frac{\pi}{2}
\end{array} \right.\label{eq:hstd}
\end{align}
where $R$ is the radius of the circular arc and $d$ is the separation
between them. The angular co-ordinate $\theta$ is in the usual
sense. The form of stadium of revolution in three dimensions in
spherical polar coordinate is simply given by
\begin{align}
r(\theta, \phi)= {\mathcal S}\left(\theta-\frac{\pi}{2}\right); \qquad ~\text{for}~~\theta \in [0,\pi]~~\text{and}~~\phi \in [0, 2\pi].\label{eq:std}
\end{align}
The shape of the stadium of revolution in three dimensions is depicted
below in fig \eqref{fig:std3}.
\subsection{Spheroids}
\begin{figure}[htb]
\centering
\subfloat[~Oblate, $r_{c}/r_{a} = 0.8$]{\includegraphics[width =
    0.25\textwidth,trim=0in 0.5in 0in 0in,clip=true]{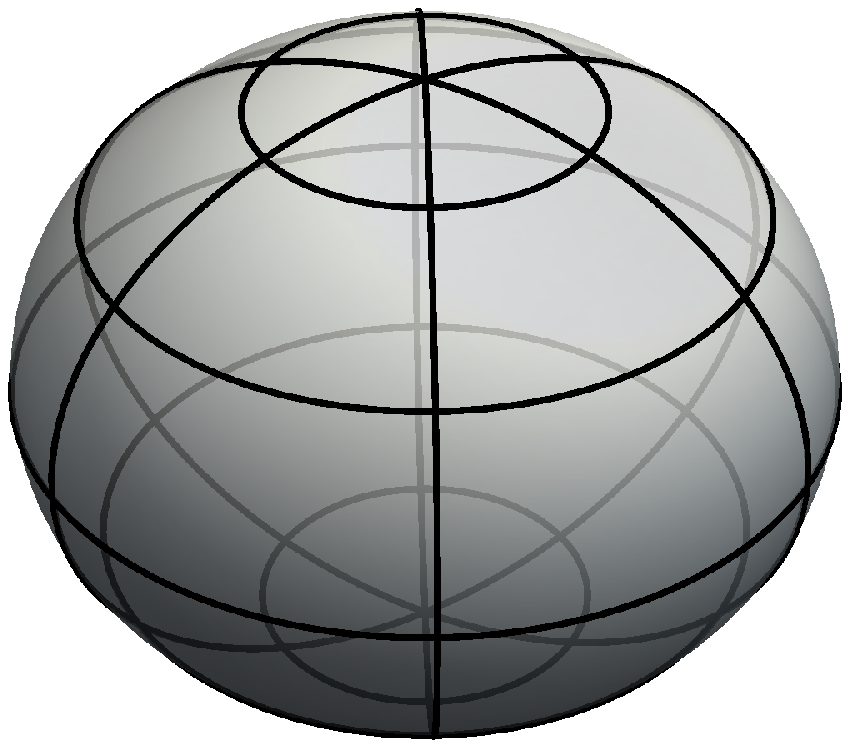}\label{fig:ob1}}
\subfloat[~Prolate, $r_{c}/r_{a} = 1.2$]{\includegraphics[width = 0.25\textwidth,trim=0in
    1.25in 0in 0.75in,clip=true]{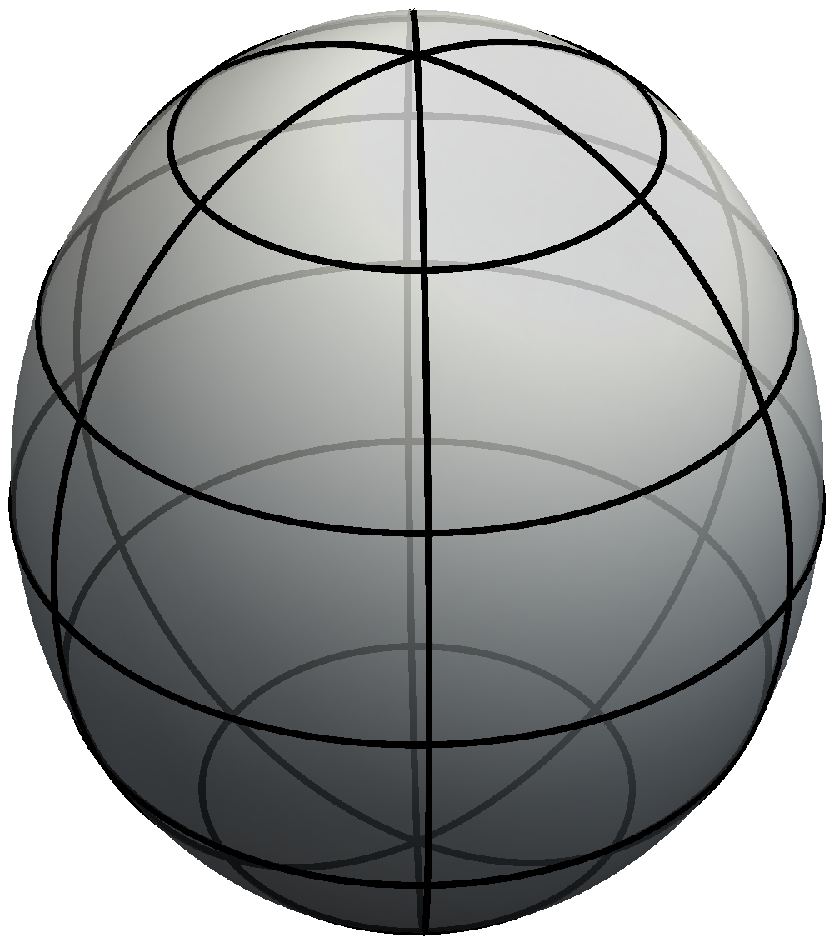}\label{fig:pro1}}
\caption{Spheroids for different ratios of polar radius ($r_{c}$) to
  equatorial radius ($r_{a}$).}
\label{fig:sph1}
\end{figure}

A spheroid in the spherical polar coordinates is given by
\begin{equation}
r(\theta, \phi) =
\frac{r_{a}}{\sqrt{1-\left[1-\left(r_{a}/r_{c}\right)^{2}\right] \cos^2
    \theta}}; \qquad  \theta \in [0,\pi]~~\text{and}~~\phi \in [0, 2\pi], \label{eq:rsph}
\end{equation}
where $r_{a}$ and $r_{c}$ $(>0)$ are called equatorial and polar radii
respectively. For $r_{c} < r_{a}$ the spheroid is known as oblate
while for $r_{c} > r_{a}$ it is called prolate. We have considered
$r_{c}/r_{a} = 0.8$ for oblate and $r_{c}/r_{a} = 1.2$ for prolate as
case studies (since these two values are on the opposite sides of the
sphere ($r_{c}= r_{a} = 1$) and have considerable deformation from it)
to show the applicability of the formalism. It is evident form
\eqref{eq:rsph} that an oblate fig \eqref{fig:ob1} (a prolate fig
\eqref{fig:pro1}) is generated by rotating an ellipse about its minor
(major) axis.
\subsection{Pear shaped enclosures}
\begin{figure}[htb]
\centering
%\subfloat[~($\beta_{2},\beta_{3},\beta_{4}$) =
%  $(0.119,0.095,0.002)$]{\includegraphics[width=0.33\textwidth,trim=0in 1.1in
%    0in 0.9in,clip=true]{./plots/pear1}\label{fig:pear1}}
%\subfloat[~($\beta_{2},\beta_{3},\beta_{4}$)
%  = $(0.154,0.097,0.080)$]{\includegraphics[width =0.33\textwidth,trim=0in 1.1in
%    0in 1.0in,clip=true]{./plots/pear2}\label{fig:pear2}}
\includegraphics[scale=0.3,trim=0in 1.0in 0in 0.45in,clip=true]{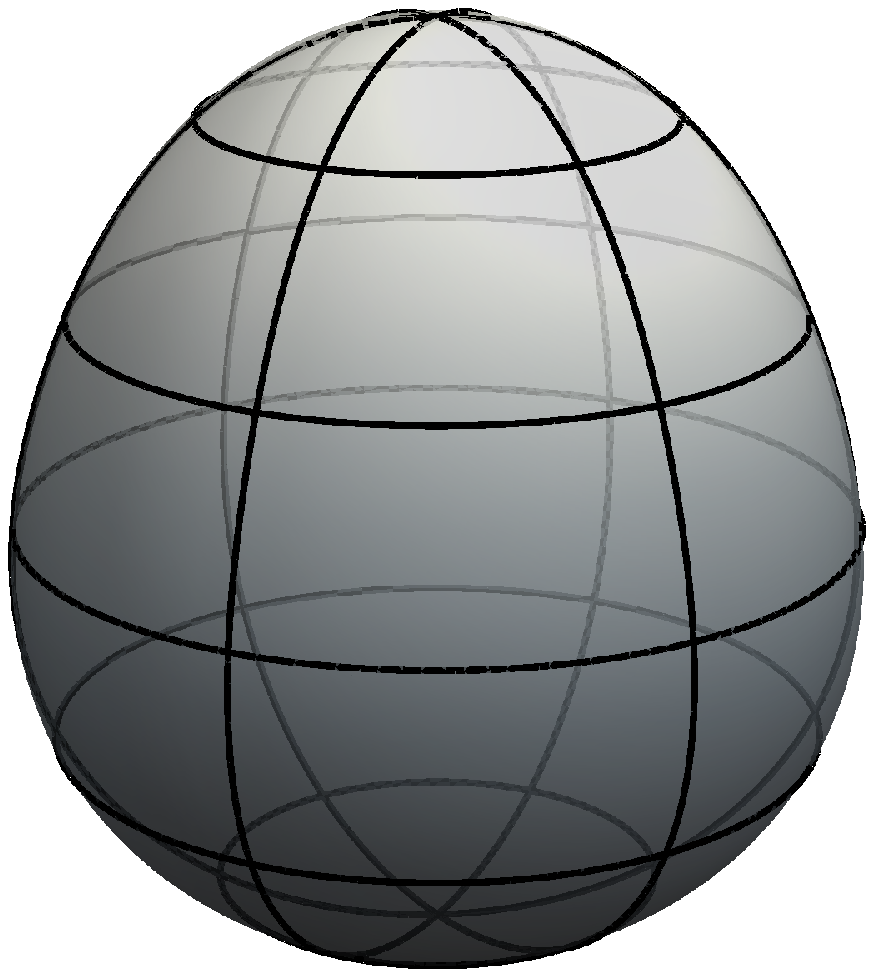}
\includegraphics[scale=0.3,trim=0in 1.0in 0in 0.45in,clip=true]{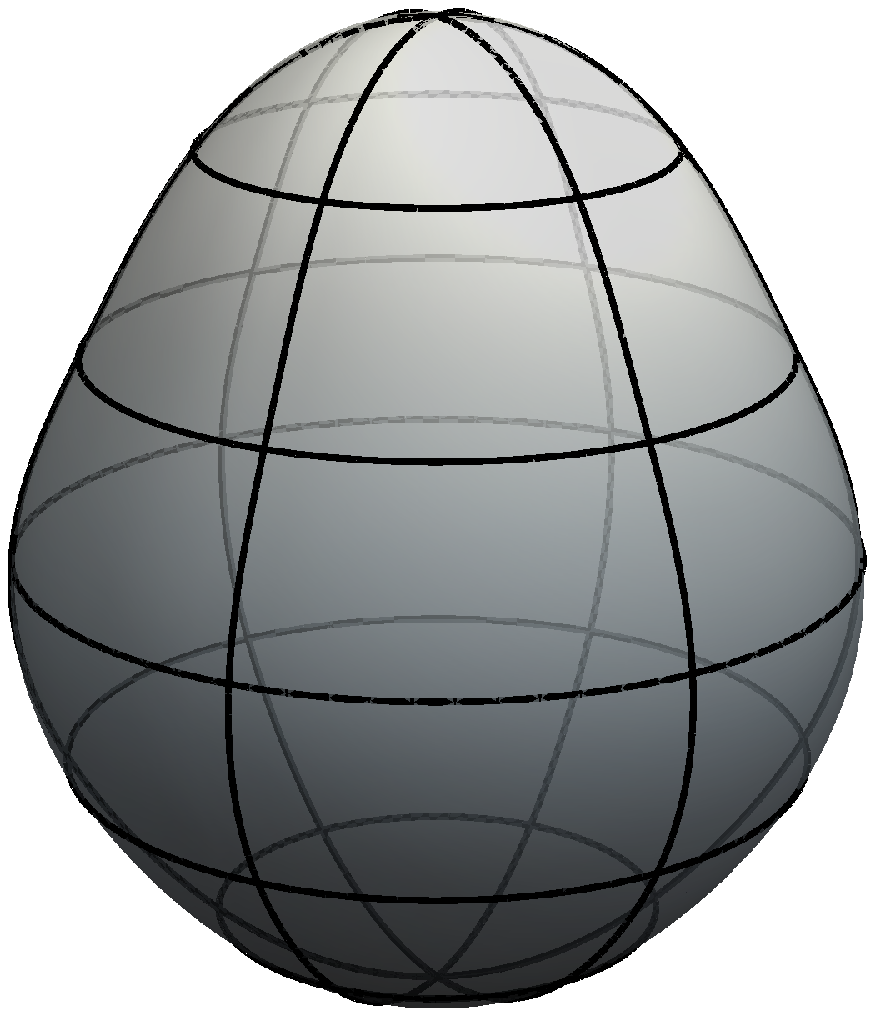}
\caption{The pear shaped domains for different parameter values. The
  figures in the left and right are for the parameter values $\{C_{2},
  C_{3}, C_{4}\}$ =$\{0.119,0.095,0.002\}$ and $\{0.154,0.097,0.080\}$
  respectively.}
\label{fig:pear}
\end{figure}

Finally, we consider the following pear shaped domains motivated by
the recent work \cite{nature13} in the experimental nuclear physics
for the $^{220}$Rn and $^{224}$Ra nuclei. The equation of the pear
shape in spherical polar coordinate is given by
\begin{align}
r(\theta, \phi) = R_{0}\left(1 + \sum \limits_{a=2}^{4} C_{a}
Y_{a}^{0}(\theta, \phi)\right); \qquad ~\text{for}~~\theta \in
[0,\pi]~~\text{and}~~\phi \in [0, 2\pi], \label{eq:pear} 
\end{align}
where $R_{0}$ is the radius of equal volume spherical shape, which we
set to one. The values for the expansion coefficients $C_{a}$s are
identified with the parameter $\beta_{a}$s and chosen as the same
given in \cite{nature13, ref10nat}. The shapes with these values are
shown in fig \eqref{fig:pear}.
\section{Results and Discussions} \label{res} 
\noindent These particular forms of $r(\theta, \phi)$, given in
\eqref{eq:rseg}, \eqref{eq:rcyld}, \eqref{eq:std}, \eqref{eq:rsph} and
\eqref{eq:pear}, have been used to estimate the metric deformation in
terms of the spherical harmonic expansion coefficients and those
coefficients have been used for the calculation of eigenfunction
corrections as well as energy corrections up to the second order of
perturbation. The specific examples chosen here are motivated by
physical problems and they possess axisymmetric property which is
required for our formalism in the cases of degenerate states. The
perturbative series converges quickly with the main contribution
coming from the first few terms. It is clear that the convergence of
the expansion coefficients will guarantee the convergence of the
perturbative series for the energy as the $m^{\rm{th}}$ order
corrections are $m$-linear in the expansion coefficients. The
following figure \eqref{fig:con} shows the convergence of the
spherical harmonics expansion coefficients for the above mentioned
shapes (for the pear shapes it is obvious from the deformation
parameters given in fig \eqref{fig:pear}). The method seems to work
quite well for the geometries having no sharp corners.
\begin{figure}[htb]
\centering
\subfloat[~Superegg]{\includegraphics[width=0.45\textwidth,trim=0in 0.0in 0in 0.0in,clip=true]{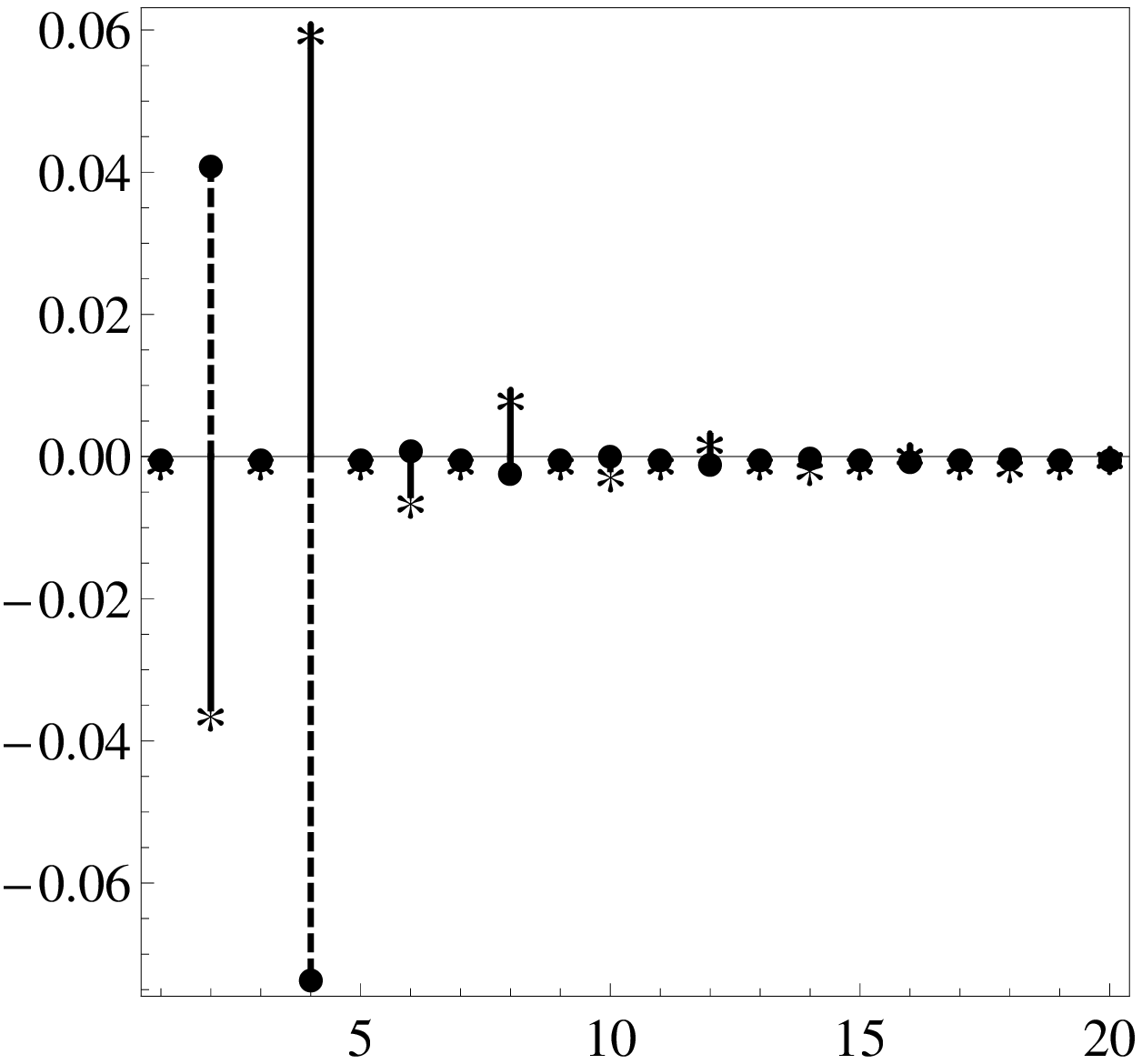}\label{fig:c1}}
\subfloat[~Stadium of revolution]{\includegraphics[width=0.45\textwidth,trim=0in 0.0in 0in 0.0in,clip=true]{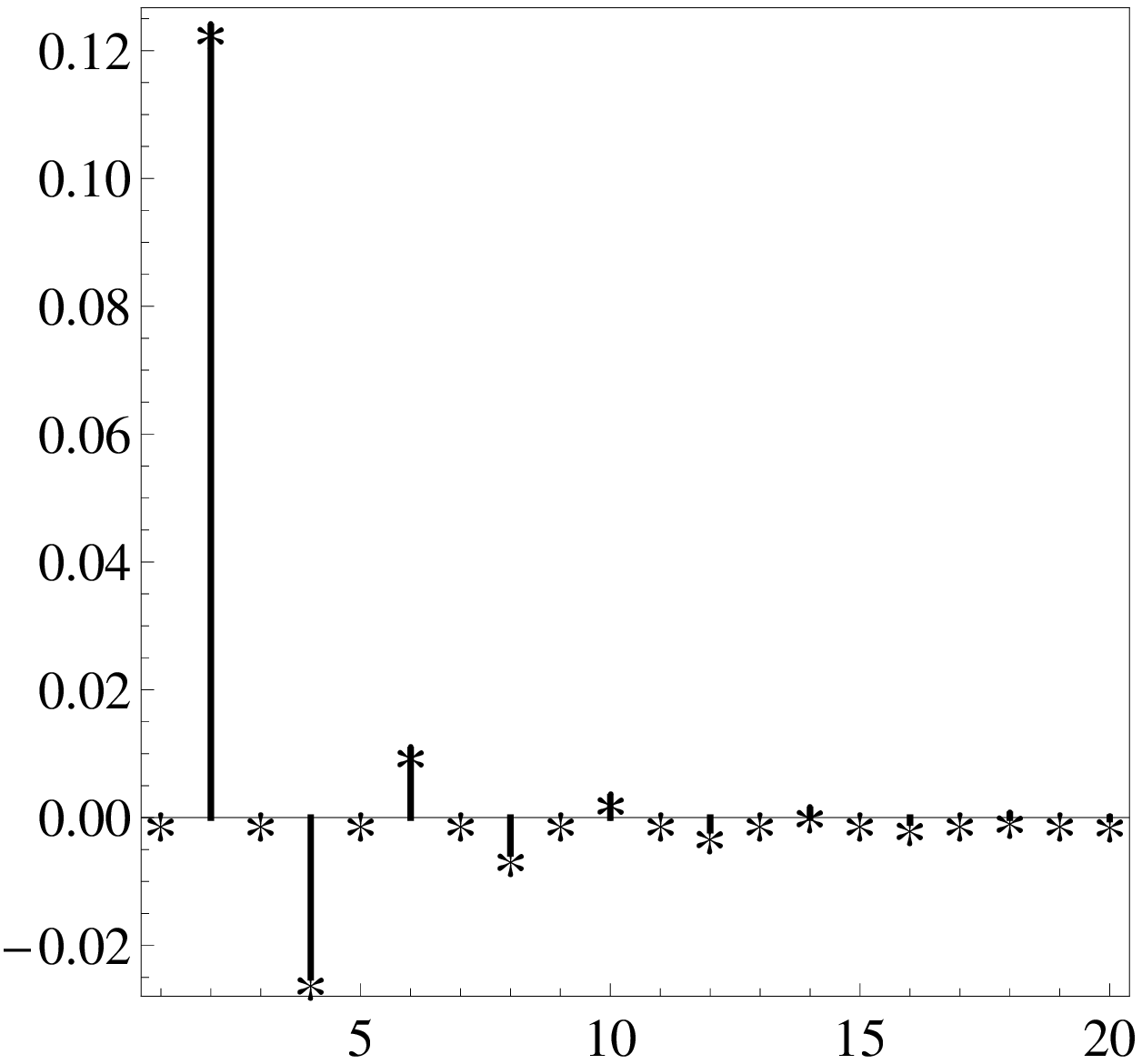}\label{fig:c2}}
\\
\subfloat[~Spheroid]{\includegraphics[width=0.45\textwidth,trim=0in 0.0in 0in
    0.0in,clip=true]{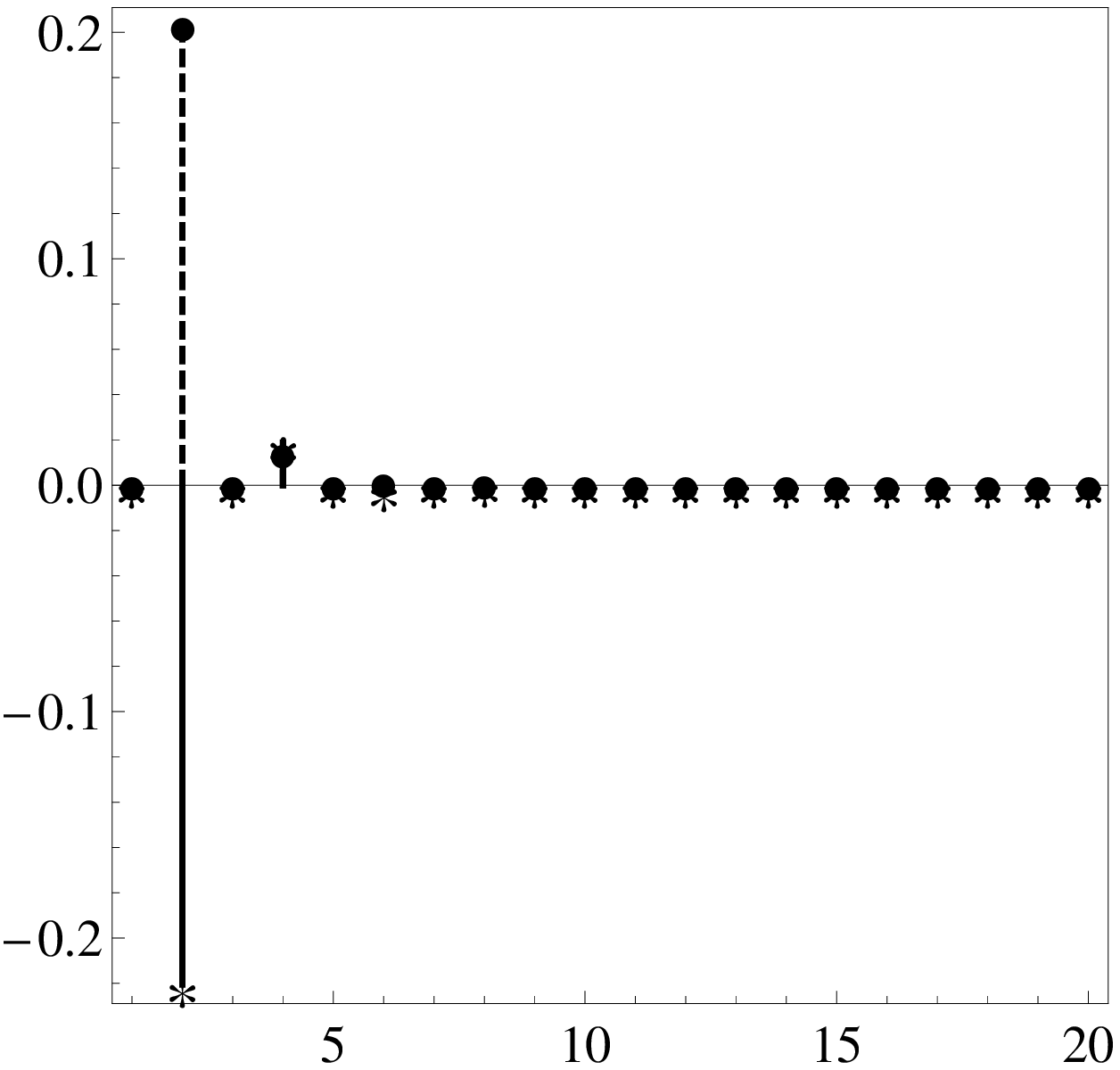}\label{fig:c3}}
\subfloat[~Rounded cylinder]{\includegraphics[width=0.45\textwidth,trim=0in 0.0in 0in 0.0in,clip=true]{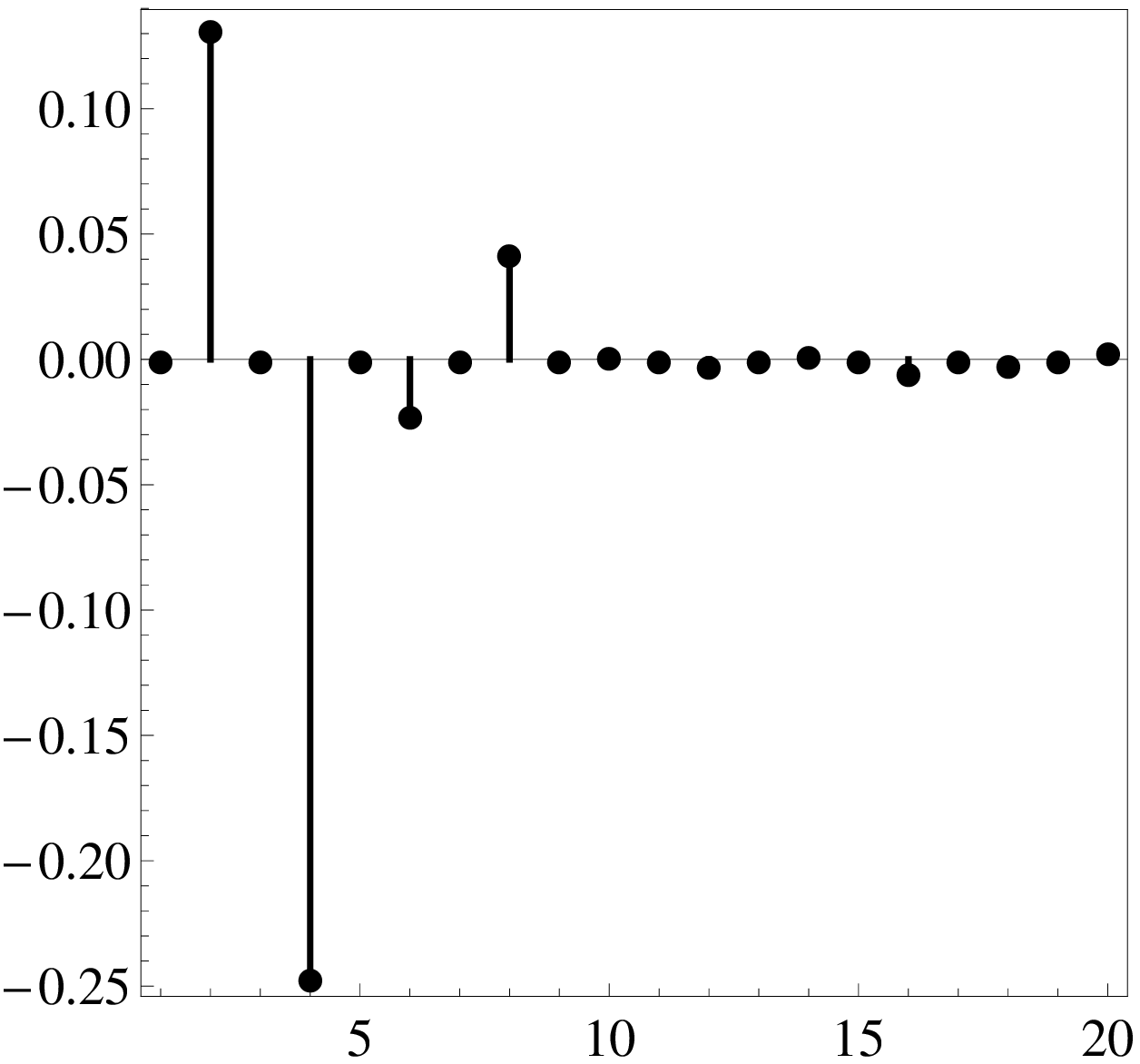}\label{fig:c4}}
\caption{This shows the convergence of the expansion coefficients
  ($C_{a}$) with the number of terms. In fig \eqref{fig:c1} solid
  (dotted) lines indicate the coefficients for a superegg with
  exponent $n=1.7$ ($n=2.5$) and similarly, in fig \eqref{fig:c3} for
  oblate (prolate).  Figs \eqref{fig:c2} and \eqref{fig:c4} show the
  coefficients for a stadium of revolution and a rounded cylindrical
  enclosure respectively.}
\label{fig:con}
\end{figure}
in this formalism, the only approximation introduced is due to the
restriction of the perturbative series to the second order. In
principle the higher order corrections could also be calculated
exactly and the results could be improved further. As it is evident
from the results that the third and higher order corrections would be
meaningful only when one has violent deformations of a sphere and
where the $C_{a}$s are not small and not converging rapidly. We have
compared the eigenvalues obtained analytically by our method with the
numerical ones for supereggs and a stadium of revolution together in
table \eqref{tab:A} and spheroids and a rounded cylinder in table
\eqref{tab:B} for both the cases of boundary conditions. In the case
of DBC we have considered energy levels (including degenerate ones) up
to the $17^{\rm{th}}$ state while for NBC those are up to the
$16^{\rm{th}}$ state. The percentage deviations of the analytical
values from the corresponding numerical ones are also shown. The
results imply that this perturbative formalism works considerably well
even for higher excited states and also for domains which are highly
deformed. For this extreme cases addition of higher order corrections
will surely improve the matching but they are algebraically
complicated. Due to the deformation from the sphere the degenerate
states, of multiplicities $(2l+1)$ for a given $l$ value, will be
splitted. It is observed that for degenerate states the energy
corrections are same for equal $|m|$ values and which implies that we
get only $(l+1)$ distinct energy values. In case of a spherical
boundary the energy levels having degeneracy $\{3\}$ or $\{5\}$
transformed after the deformation into $\{2,1\}/\{1,2\}$ or
$\{1,2,2\}/\{2,2,1\}/\{2,1,2\}$-fold degeneracy respectively. Also the
energy levels for different $l,m$ values are interlaced among each
other after the deformation in a wild fashion for higher excited
states and due to that we restrict the comparison only for low-lying
levels. The matching between the analytical results and the
corresponding numerical counterparts are excellent for the shapes
which are not highly deformed and for non-trivial shapes also the
agreement is quite satisfactory. In case of DBC, in table
\eqref{tab:A} the discrepancy is $< 0.01\%$ for supereggs and stadium
of revolution with a maximum of $\sim 0.7\%$ while for NBC it is
relatively larger ($< 0.1\%$) with a maximum of $\sim 0.2\%$ for the
stadium. Next we have clubbed oblate, prolate and rounded cylinder
domains together. Even though the magnitude of deviation from the
sphere is same for the oblate and prolate considered here, the shapes
are completely different and as a result the matching is also
dissimilar. In table \eqref{tab:B} the mismatch is $< 1\%$ and maximum
deviations for the oblate is $\sim 1.5\%$ and for the prolate it is
mostly $< 1.5\%$ and goes up to $\sim 4\%$ for couple of levels. For
rounded cylinder the errors are within $\sim 1.7\%$. In case of NBC
for the same group of boundaries the matching is very good (with
errors are $\sim 1\%$ for spheroids and $< 2\%$ for rounded cylinder)
and the maximum discrepancies are $\sim 1.6\%$, $\sim 2.5\%$ and $\sim
3.9\%$ respectively for oblate, prolate and rounded cylinder
enclosures for a few specific levels. Finally in the last table
\eqref{tab:C} we have compared the results for pear shaped domains
satisfying both the DBC and the NBC with an analogy with the nuclear
physics. The agreement between the analytic perturbative results and
the numerical ones is better for the pear shape for which the $C_{a}$
values are small compared to other. The maximum deviation goes up to
$2\%$ in the worst case for the DBC. While for the NBC it is evident
from table \eqref{tab:C} that the matching is excellent barring a
couple of cases marked with stars ($\star$) and daggers ($\dagger$)
(where it is disastrous) in that table. For these cases the mismatch
is solely due to the accidental closeness of the zeros of derivative
of spherical Bessel functions $j_{0}$ and $j_{3}$, which are
$\alpha_{1,0} (=4.49341)$ and $\alpha_{1,3} (=4.51409)$
respectively. In the calculation of the second order energy correction
for $l=0$ due to the non-zero finite value of $C_{3}$, the term of the
form, $-\left(1+
\frac{\alpha_{1,0}j_{3}(\alpha_{1,0})}{j^{\prime}_{3}(\alpha_{1,0})}\right)\frac{C^{2}_{3}}{2\pi}$,
picks up a very high negative value (as the denominator is very close
to zero) and causes the problem in the levels marked by stars
($\star$) for both the shapes. Similarly, for $l=3$ and $m=0$ the
second order energy correction term, containing
$\frac{\alpha_{1,3}j_{0}(\alpha_{1,3})}{j^{\prime}_{0}(\alpha_{1,3})}$,
produces a large positive correction as the denominator is again very
close to zero and which creates the error for the levels marked by
daggers ($\dagger$) for both the cases. Other than the above mentioned
exception the matching between the perturbative and numerical values
are outstanding. The agreement between the numerical and perturbative
values are better for supereggs than that for spheroids for both the
cases of boundary conditions and is analogous to the behaviour of
their two dimensional counterparts i.e. between supercircle and
ellipse as reported in earlier works \cite{p1ref3,p2ref33}. We would
like to point out that the eigenvalue correction for the state $l=2,
m=0$ behaves in a peculiar fashion and disrupts the matching between
different levels and produces the instances of maximum deviations for
most of the cases in the above mentioned examples for both the
boundary conditions. This abnormality for $l=2$ is an issue which
calls for further investigations. Another unique feature of this
method is that it can capture the degeneracy patterns for apparently
similar looking shapes having different degeneracy patterns and also
distinct shapes having similar type of degeneracies. As an example,
the prolate, the stadium of revolution and the superegg ($n=2.5$) are
looking similar apparently but the first two shapes have similar
degeneracy pattern only for the DBC. The distinction is made in their
NBC degeneracy patterns. In contrast, rounded cylinder and superegg
($n=2.5$) are having different structures but they produce similar
degeneracy pattern for the DBC. So, our perturbative method traces the
degeneracy patterns (provided by the numerical results) in a correct
fashion for low-lying levels for different shapes and both the cases
of boundary conditions. On the other hand, there are a few instances
where three very close energylevels show degeneracy mismatch
(viz. superegg ($n=1.7$) for DBC levels $5, 6$ and $7$; oblate for NBC
levels $13, 14$ and $15$; rounded cylinder for NBC levels $11, 12$ and
$13$). We expect that inclusion of higher order correction term may
provide a positive correction to the non-degenerate state whereas the
two degenerate states may get a negative correction and which might
bring them in order.

In conclusion, we point out that expansion of the boundary asymmetry
in terms of spherical harmonics makes the method completely general to
encompass a large variety of domains in three dimensions. The closed
form corrections for wavefunctions and energies at each order of
perturbation in terms of expansion coefficients help us to apply the
prescription for a general boundary conditions also. Since the
solutions are boundary condition free this method can tackle variety
of problems. Also the boundary conditions maintain a simple form at
each order of perturbation at the cost of complicating the
equations. This method can treat both degenerate and non-degenerate
states in the same way for axisymmetric boundaries. This method has an
edge over the other method discussed by Panda and Hazra recently
\cite{p5ref7}. In the present method applying boundary condition is
easy in contrast to the other method where it is extremely
cumbersome. In this case all order equations as well as energy
corrections are also written.

\section*{Acknowledgements}
SP would like to acknowledge the Council of Scientific and Industrial
Research (CSIR), India for providing the financial support.

\begin{table}[tp]\scriptsize
\caption{Comparison of the low-lying energy eigenvalues (including
  degeneracies) with the magnitude of $\%$ error (=
  $\rvert\frac{Ns-Ps}{Ns}\lvert \times 100\%$) for superegg with
  exponent $n = 1.7$ and $2.5$ and stadium of revolution.}  \setlength{\tabcolsep}{8pt}
\renewcommand{\arraystretch}{1.0}
\label{tab:A}
\begin{tabular}{*{9}{c}}\toprule
\multicolumn{3}{c}{Superegg ($n=1.7$)}&\multicolumn{3}{c}{Superegg ($n=2.5$)}&\multicolumn{3}{c}{Stadium}\\\colrule
$Ps$& $Ns$ &$\%$ Error & $Ps$ & $Ns$ &$\%$ Error & $Ps$ & $Ns$ &$\%$
Error \\\toprule
\multicolumn{9}{c}{\bf Dirichlet Boundary Condition} \\ \toprule
 10.670 & 10.669 & 0.009 & 9.169 & 9.169 & 0.000 & 8.857 & 8.857 & 0.000 \\
 21.618 & 21.617 & 0.005 & 18.355 & 18.355 & 0.000 & 17.002 & 16.998 & 0.024 \\
 21.618 & 21.617 & 0.005 & 18.929 & 18.930 & 0.005 & 18.669 & 18.670 & 0.005 \\
 22.198 & 22.198 & 0.000 & 18.929 & 18.930 & 0.005 & 18.669 & 18.670 & 0.005 \\
 35.239 & 35.239 & 0.000 & 29.873 & 29.871 & 0.007 & 28.074 & 28.274 & 0.707 \\
 35.240 & 35.239 & 0.003 & 29.873 & 29.871 & 0.007 & 28.848 & 28.846 & 0.007 \\
 35.240 & 35.246 & 0.017 & 31.385 & 31.373 & 0.038 & 28.848 & 28.846 & 0.007 \\
 36.795 & 36.796 & 0.003 & 31.439 & 31.439 & 0.000 & 31.171 & 31.172 & 0.003 \\
 36.795 & 36.796 & 0.003 & 31.439 & 31.439 & 0.000 & 31.171 & 31.172 & 0.003 \\
 42.644 & 42.636 & 0.019 & 36.627 & 36.644 & 0.046 & 36.257 & 36.047 & 0.583 \\
 51.439 & 51.440 & 0.002 & 44.087 & 44.085 & 0.005 & 41.634 & 41.640 & 0.014 \\
 51.439 & 51.440 & 0.002 & 44.087 & 44.085 & 0.005 & 42.018 & 42.184 & 0.394 \\
 52.499 & 52.501 & 0.004 & 44.753 & 44.781 & 0.063 & 42.018 & 42.184 & 0.394 \\
 52.722 & 52.712 & 0.019 & 44.753 & 44.781 & 0.063 & 43.231 & 43.229 & 0.005 \\
 52.722 & 52.712 & 0.019 & 45.640 & 45.634 & 0.013 & 43.231 & 43.229 & 0.005 \\
 53.892 & 53.894 & 0.004 & 46.556 & 46.554 & 0.004 & 46.245 & 46.246 & 0.002 \\
 53.892 & 53.894 & 0.004 & 46.556 & 46.554 & 0.004 & 46.245 & 46.246 & 0.002 \\ \toprule
\multicolumn{9}{c}{\bf Neumann Boundary Condition} \\ \toprule
 4.561 & 4.562 & 0.022 & 3.794 & 3.796 & 0.053 & 3.345 & 3.345 & 0.000 \\
 4.561 & 4.562 & 0.022 & 4.088 & 4.088 & 0.000 & 4.145 & 4.143 & 0.048 \\
 4.842 & 4.841 & 0.021 & 4.088 & 4.088 & 0.000 & 4.145 & 4.143 & 0.048 \\
 11.233 & 11.237 & 0.036 & 9.508 & 9.510 & 0.021 & 9.387 & 9.389 & 0.021 \\
 11.594 & 11.596 & 0.017 & 9.508 & 9.510 & 0.021 & 9.387 & 9.389 & 0.021 \\
 11.594 & 11.596 & 0.017 & 10.724 & 10.720 & 0.037 & 9.515 & 9.535 & 0.209 \\
 12.855 & 12.855 & 0.000 & 10.724 & 10.720 & 0.037 & 10.747 & 10.742 & 0.047 \\
 12.855 & 12.855 & 0.000 & 11.144 & 11.136 & 0.072 & 10.747 & 10.742 & 0.047 \\
 21.014 & 21.020 & 0.029 & 17.751 & 17.755 & 0.023 & 17.089 & 17.099 & 0.058 \\
 21.014 & 21.020 & 0.029 & 17.751 & 17.755 & 0.023 & 17.591 & 17.620 & 0.165 \\
 21.566 & 21.566 & 0.000 & 18.522 & 18.545 & 0.124 & 17.591 & 17.620 & 0.165 \\
 21.637 & 21.623 & 0.065 & 18.710 & 18.714 & 0.021 & 17.716 & 17.719 & 0.017 \\
 21.882 & 21.879 & 0.014 & 18.710 & 18.714 & 0.021 & 17.716 & 17.719 & 0.017 \\
 21.882 & 21.879 & 0.014 & 19.351 & 19.349 & 0.010 & 18.279 & 18.241 & 0.208 \\
 23.070 & 23.068 & 0.009 & 19.723 & 19.713 & 0.051 & 19.687 & 19.677 & 0.051 \\
 23.070 & 23.068 & 0.009 & 19.723 & 19.713 & 0.051 & 19.687 & 19.677 & 0.051 \\  \bottomrule
\end{tabular}

\end{table}

\begin{table}[tp]\scriptsize
\caption{Comparison of the low-lying energy eigenvalues (including
  degeneracies) with the magnitude of $\%$ error (=
  $\rvert\frac{Ns-Ps}{Ns}\lvert \times 100\%$) for oblate
  ($r_{c}/r_{a} = 0.8$), prolate ($r_{c}/r_{a} = 1.2$) and rounded
  cylinder.}  \setlength{\tabcolsep}{8pt}
\renewcommand{\arraystretch}{1.0}
\label{tab:B}
\begin{tabular}{*{9}{c}}\toprule
\multicolumn{3}{c}{Oblate}&\multicolumn{3}{c}{Prolate}&\multicolumn{3}{c}{Rounded
cylinder}\\\colrule
$Ps$& $Ns$ &$\%$ Error & $Ps$ & $Ns$ &$\%$ Error & $Ps$ & $Ns$ &$\%$
Error \\\toprule
\multicolumn{9}{c}{\bf Dirichlet Boundary Condition} \\ \toprule
 10.060 & 10.081 & 0.208 & 10.013 & 9.998 & 0.150 & 11.104 & 11.079 & 0.226 \\
 19.383 & 19.314 & 0.357 & 18.517 & 18.580 & 0.339 & 21.243 & 21.154 & 0.421 \\
 19.383 & 19.314 & 0.357 & 21.452 & 21.379 & 0.342 & 23.057 & 23.044 & 0.056 \\
 22.936 & 23.201 & 1.142 & 21.452 & 21.379 & 0.342 & 23.057 & 23.044 & 0.056 \\
 31.040 & 30.857 & 0.593 & 28.845 & 30.122 & 4.239 & 33.568 & 33.393 & 0.524 \\
 31.040 & 30.857 & 0.593 & 32.475 & 32.512 & 0.114 & 33.568 & 33.393 & 0.524 \\
 33.519 & 33.324 & 0.585 & 32.475 & 32.512 & 0.114 & 38.381 & 37.847 & 1.411 \\
 35.220 & 35.380 & 0.452 & 35.986 & 35.829 & 0.438 & 38.788 & 38.746 & 0.108 \\
 35.220 & 35.380 & 0.452 & 35.986 & 35.829 & 0.438 & 38.788 & 38.746 & 0.108 \\
 43.269 & 43.931 & 1.507 & 42.603 & 41.306 & 3.139 & 43.617 & 44.246 & 1.422 \\
 44.937 & 44.616 & 0.720 & 43.648 & 44.427 & 1.753 & 49.331 & 49.401 & 0.142 \\
 44.937 & 44.616 & 0.720 & 45.636 & 46.308 & 1.451 & 49.331 & 49.401 & 0.142 \\
 49.697 & 49.398 & 0.605 & 45.636 & 46.308 & 1.451 & 49.571 & 50.458 & 1.758 \\
 49.697 & 49.398 & 0.605 & 49.421 & 49.391 & 0.061 & 49.571 & 50.458 & 1.758 \\
 50.421 & 49.748 & 1.352 & 49.421 & 49.391 & 0.061 & 55.717 & 54.790 & 1.692 \\
 50.421 & 49.748 & 1.352 & 53.473 & 53.210 & 0.494 & 58.114 & 57.969 & 0.250 \\
 51.056 & 51.533 & 0.926 & 53.473 & 53.210 & 0.494 & 58.114 & 57.969 & 0.250 \\\toprule
\multicolumn{9}{c}{\bf Neumann Boundary Condition} \\ \toprule
 3.809 & 3.775 & 0.901 & 3.399 & 3.460 & 1.763 & 3.593 & 3.706 & 3.049 \\
 3.809 & 3.775 & 0.901 & 4.870 & 4.836 & 0.703 & 4.782 & 4.771 & 0.231 \\
 5.571 & 5.662 & 1.607 & 4.870 & 4.836 & 0.703 & 4.782 & 4.771 & 0.231 \\
 9.839 & 9.750 & 0.905 & 9.419 & 9.663 & 2.525 & 8.835 & 8.926 & 1.019 \\
 9.839 & 9.750 & 0.905 & 10.648 & 10.693 & 0.421 & 8.835 & 8.926 & 1.019 \\
 12.003 & 11.900 & 0.866 & 10.648 & 10.693 & 0.421 & 13.185 & 13.011 & 1.337 \\
 12.104 & 12.145 & 0.338 & 12.539 & 12.453 & 0.691 & 13.185 & 13.011 & 1.337 \\
 12.104 & 12.145 & 0.338 & 12.539 & 12.453 & 0.691 & 14.795 & 14.259 & 3.759 \\
 17.954 & 17.788 & 0.933 & 17.918 & 18.182 & 1.452 & 17.422 & 17.696 & 1.548 \\
 17.954 & 17.788 & 0.933 & 18.731 & 18.939 & 1.098 & 17.422 & 17.696 & 1.548 \\
 20.683 & 20.678 & 0.024 & 18.731 & 18.939 & 1.098 & 19.490 & 20.076 & 2.919 \\
 20.683 & 20.678 & 0.024 & 20.564 & 20.561 & 0.015 & 19.831 & 20.076 & 1.220 \\
 21.260 & 21.449 & 0.881 & 20.564 & 20.561 & 0.015 & 19.831 & 20.637 & 3.906 \\
 21.601 & 21.449 & 0.709 & 21.061 & 20.850 & 1.012 & 24.559 & 24.166 & 1.626 \\
 21.601 & 21.581 & 0.093 & 22.887 & 22.733 & 0.677 & 24.891 & 24.436 & 1.862 \\
 22.010 & 22.065 & 0.249 & 22.887 & 22.733 & 0.677 & 24.891 & 24.436 & 1.862 \\
 \bottomrule
\end{tabular}

\end{table}

\begin{table}[tp]\scriptsize
\caption{Comparison of the low-lying energy eigenvalues (including
  degeneracies) with the magnitude of $\%$ error (=
  $\rvert\frac{Ns-Ps}{Ns}\lvert \times 100\%$) for the pear shapes
  with the parameters $(C_{2}, C_{3}, C_{4})$ = $(0.119,0.095,0.002)$
  and $(0.154,0.097,0.080)$ respectively.}
\setlength{\tabcolsep}{8pt} \renewcommand{\arraystretch}{1.0}
\label{tab:C}
\begin{tabular}{*{6}{c}}\toprule
\multicolumn{6}{c}{Pear shapes}\\ \colrule
\multicolumn{3}{c}{($C_{2}, C_{3}, C_{4}$) =
  $(0.119,0.095,0.002)$}&\multicolumn{3}{c}{($C_{2}, C_{3}, C_{4}$)
  = $(0.154,0.097,0.080)$}\\\colrule
$Ps$& $Ns$ &$\%$ Error & $Ps$ & $Ns$ &$\%$ Error\\\toprule
\multicolumn{6}{c}{\bf Dirichlet Boundary Condition} \\ \toprule
 9.938 & 9.934 & 0.040 & 9.999 & 9.982 & 0.169 \\
 19.088 & 19.085 & 0.016 & 18.938 & 18.900 & 0.199 \\
 20.858 & 20.852 & 0.029 & 21.078 & 21.063 & 0.072 \\
 20.858 & 20.852 & 0.029 & 21.078 & 21.063 & 0.072 \\
 31.216 & 31.369 & 0.488 & 29.403 & 29.874 & 1.577 \\
 32.667 & 32.672 & 0.015 & 33.405 & 33.403 & 0.004 \\
 32.667 & 32.672 & 0.015 & 33.405 & 33.403 & 0.004 \\
 34.669 & 34.651 & 0.052 & 34.913 & 34.893 & 0.058 \\
 34.669 & 34.651 & 0.052 & 34.913 & 34.893 & 0.058 \\
 40.027 & 39.841 & 0.467 & 40.651 & 39.848 & 2.015 \\
 46.964 & 46.815 & 0.318 & 43.765 & 44.386 & 1.401 \\
 47.090 & 47.176 & 0.182 & 46.830 & 46.851 & 0.044 \\
 47.090 & 47.176 & 0.182 & 46.830 & 46.851 & 0.044 \\
 49.051 & 49.071 & 0.041 & 50.539 & 50.554 & 0.030 \\
 49.051 & 49.071 & 0.041 & 50.539 & 50.554 & 0.030 \\
 51.242 & 51.202 & 0.078 & 51.398 & 51.372 & 0.052 \\
 51.242 & 51.202 & 0.078 & 51.398 & 51.372 & 0.052 \\\toprule
\multicolumn{6}{c}{\bf Neumann Boundary Condition} \\ \toprule
 3.681 & 3.694 & 0.352 & 3.445 & 3.466 & 0.606 \\
 4.601 & 4.597 & 0.087 & 4.684 & 4.682 & 0.043 \\
 4.601 & 4.597 & 0.087 & 4.684 & 4.682 & 0.043 \\
 10.319 & 10.334 & 0.145 & 8.833 & 8.982 & 1.659 \\
 10.798 & 10.796 & 0.019 & 11.444 & 11.416 & 0.245 \\
 10.798 & 10.796 & 0.019 & 11.444 & 11.416 & 0.245 \\
 11.853 & 11.839 & 0.118 & 11.863 & 11.861 & 0.017 \\
 11.853 & 11.839 & 0.118 & 11.863 & 11.861 & 0.017 \\
 5.072 & 18.228 & $\star$ & 4.344 & 16.927 & $\star$ \\
 19.358 & 19.380 & 0.114 & 18.820 & 18.847 & 0.143 \\
 19.358 & 19.380 & 0.114 & 18.820 & 18.847 & 0.143 \\
 20.439 & 20.450 & 0.054 & 33.482 & 21.016 & $\dagger$ \\
 20.439 & 20.450 & 0.054 & 21.703 & 21.485 & 1.015 \\
 34.592 & 21.377 & $\dagger$ & 21.703 & 21.485 & 1.015 \\
 21.630 & 21.597 & 0.153 & 21.487 & 21.680 & 0.898 \\
 21.630 & 21.597 & 0.153 & 21.487 & 21.680 & 0.898 \\
 \bottomrule
\end{tabular}

\end{table}

\bibliography{refbibtex}

\end{document}